\documentclass[aip, pof, reprint]{revtex4-1}
\usepackage{graphicx}
\usepackage{color}
\usepackage{amssymb}
\usepackage{amsmath}
\usepackage{tabularx}
\usepackage{bm}
\usepackage{hyperref}
\usepackage{ltxtable}
\usepackage{natbib} 
\usepackage{longtable}
\usepackage{natbib}

\usepackage[usenames,dvipsnames]{xcolor}
\hypersetup{
colorlinks,
citecolor=blue,
filecolor=blue,
linkcolor=blue,
urlcolor=blue}

\newcommand{\be}{\begin{equation}}
\newcommand{\ee}{\end{equation}} 
\newcommand{\bea}{\begin{eqnarray}}
\newcommand{\eea}{\end{eqnarray}}

\begin{document}

\title{Scaling and spatial intermittency of thermal dissipation in turbulent convection}

\author{Shashwat Bhattacharya}
\email{shabhatt@iitk.ac.in}
\affiliation{Department of Mechanical Engineering, Indian Institute of Technology Kanpur, Kanpur 208016, India}
\author{Ravi Samtaney}
\affiliation{Mechanical Engineering, Division of Physical Science and Engineering, King Abdullah University of Science and Technology, Thuwal 23955, Saudi Arabia}
\author{Mahendra K. Verma}
\affiliation{Department of Physics, Indian Institute of Technology Kanpur, Kanpur 208016, India}

\date{17 June 2019}%

\begin{abstract}
We derive scaling relations for the thermal dissipation rate in the bulk and in the boundary layers for moderate and large Prandtl number (Pr) convection. Using direct numerical simulations of Rayleigh-B\'{e}nard convection, we show that the thermal dissipation in the bulk is suppressed compared to passive scalar dissipation. The suppression is stronger for large Pr. We further show that the dissipation in the boundary layers dominates that in the bulk for both moderate and large \textcolor{black}{Pr}. The probability distribution functions (PDFs) of thermal dissipation rate, both in the bulk and in the boundary layers, are stretched exponential, similar to passive scalar dissipation.
\end{abstract}
\maketitle

\section{Introduction} \label{sec:Introduction}
Scalar fields, such as temperature and concentration, are often carried along by turbulent flows. Flows with scalars are ubiquitous and frequently encountered in engineering and atmospheric applications. In general, these scalar fields influence the dynamics of fluid flow. The resulting coupling between the momentum and the scalar equations, along with strong nonlinearities, makes such flows very complex. \citet{Obukhov:1949} and \citet{Corrsin:JAP1951} described the energetics of a simplified system consisting of homogeneous isotropic turbulence with passive scalar fields; such scalars do not affect the velocity field. In passive scalar turbulence, both kinetic energy [defined as $(1/2) \langle |\mathbf{u}|^2 \rangle$] and scalar energy [defined as $(1/2) \langle \theta^2 \rangle$] are supplied at large scales. Here, $\theta$ and $\mathbf{u}$ are scalar and velocity fields respectively, and $\langle \, \rangle$ denotes volume average. The supplied kinetic and scalar energies cascade to intermediate scales and then to dissipative scales. Similar to kinetic energy in homogeneous turbulence, the rate of scalar energy supply equals the scalar energy cascade rate $\Pi_\theta$ and the scalar dissipation rate $\epsilon_\theta$~\cite{Lesieur:book:Turbulence,Verma:book:BDF}. Dimensional analysis gives $\epsilon_\theta \approx U \Theta^2/L$, where $L$, $U$, and $\Theta$ are large-scale length, velocity, and scalar respectively.

In the present work, \textcolor{black}{we consider} turbulence in buoyancy-driven convection, which is an example of active scalar turbulence where the scalar field (temperature) influences the flow-dynamics. We focus on an idealized system called {\em Rayleigh--B\'{e}nard convection} (RBC) in which a fluid is enclosed between \textcolor{black}{two horizontal walls, with the bottom wall being hotter than the top one.~\cite{Ahlers:RMP2009,Lohse:ARFM2010,Verma:NJP2017}.} Each horizontal wall is isothermal. RBC is specified by two nondimensional parameters---\textcolor{black}{Rayleigh number ($\mathrm{Ra}$) and Prandtl number ($\mathrm{Pr}$). These parameters are defined as 
\begin{equation}
\mathrm{Ra} = \frac{\alpha g \Delta d^3}{\nu \kappa}, \quad \mathrm{Pr}=\frac{\nu}{\kappa}, \nonumber
\end{equation}
where $\alpha$, $\nu$, and $\kappa$ respectively are the thermal expansion coefficient, kinematic viscosity, and thermal diffusivity of the fluid, $g$ is the gravitational acceleration, and $\Delta$ and $d$ respectively are the temperature difference and the distance between the top and bottom plates.}  

The energetics of \textcolor{black}{thermally-driven} convection is more complex than that of passive scalar turbulence; this is due to the two-way coupling between the governing equations of momentum and \textcolor{black}{thermal energy} (see Sec.~\ref{sec:Equations}), along with the \textcolor{black}{presence of thermal boundary layers. Presently, we focus} on the properties of thermal dissipation rate $\epsilon_T(\mathbf{r}) = \kappa (\nabla T)^2$, where $T$ is the temperature field. In RBC, the volume-averaged thermal dissipation rate is related to the Nusselt number ($\mathrm{Nu}$) by the following relation derived by \citet{Shraiman:PRA1990}:
\begin{equation}
\epsilon_T = \left\langle \kappa (\nabla T)^2 \right\rangle  = \frac{\kappa \Delta^2}{d^2} \mathrm{Nu} = \frac{U \Delta^2}{d} \mathrm{\frac{Nu}{Pe}}.
\label{eq:SS_exact}
\end{equation}
The Nusselt number is the ratio of the total heat flux and the conductive heat flux, and $\mathrm{Pe}=Ud/\kappa$ is the P\'{e}clet number. When the thermal boundary layers are less significant than the bulk (as in the ultimate regime proposed by Kraichnan~\cite{Kraichnan:PF1962Convection}), or absent (as in a periodic box\cite{Verma:PRE2012}), \textcolor{black}{both Nu and Pe are proportional to $\sqrt{\mathrm{RaPr}}$~(See Refs.\cite{Grossmann:JFM2000,Grossmann:PRL2001,Verma:NJP2017}).} These relations, when substituted in Eq.~(\ref{eq:SS_exact}), yield $\epsilon_T \sim U \Delta^2/d$, similar to passive scalar turbulence.  

In  RBC, the thermal boundary layers near the conducting walls play an important role in the scaling of thermal dissipation rate. In our present work, \textcolor{black}{we focus} on the $\mathrm{Ra}$ dependence of thermal dissipation rate and other quantities. For moderate Prandtl numbers (\textcolor{black}{of order 1}), it has been shown via scaling arguments~\cite{Malkus:PRSA1954,Castaing:JFM1989,Grossmann:JFM2000,Grossmann:PRE2002},  experiments~\cite{Castaing:JFM1989,Qiu:PRE2002,Qiu:PF2004,Brown:JSM2007,Funfschilling:JFM2005,Nikolaenko:JFM2005,He:NJP2012,Ahlers:NJP2012, Vial:PF2017}, and numerical simulations~\cite{Verzicco:JFM2003, Scheel:JFM2012, Scheel:JFM2014, Waleffe:PF2015, Verma:PF2015Reversal, Zhou:PF2018, Pandey:PF2016, Pandey:PRE2016} that
\begin{equation}
\mathrm{Pe} \sim \mathrm{Ra}^{0.5}, \quad \mathrm{Nu} \sim \mathrm{Ra}^{0.3}. \nonumber 
\end{equation} 
Note that the exponents in the above expressions shown here are approximate. Substitution of these expressions in Eq.~(\ref{eq:SS_exact}) yields 
\be
\epsilon_T  \sim \frac{\kappa \Delta^2}{d^2} \mathrm{Ra}^{0.3} \sim \frac{U \Delta^2}{d}  \mathrm{Ra}^{-0.2}.
\label{eq:epsilon_T_total}
\ee
\textcolor{black}{When compared to passive scalar flow, the additional term $\mathrm{Ra}^{-0.2}$ in RBC accounts for suppression of nonlinear interactions due to the presence of walls}; \citet{Pandey:PF2016} and \citet{Pandey:PRE2016} showed that in RBC, the ratio of the non-linear term to the diffusive term in the \textcolor{black}{equation for thermal energy} is \textcolor{black}{proportional} to $\mathrm{PeRa}^{-0.30}$ instead of Pe. \textcolor{black}{The walls truncate some of the Fourier modes, resulting in several channels of nonlinear interations and energy cascades to be blocked~(See Ref.~\cite{Verma:book:BDF} for details). Consequently, thermal dissipation in RBC is weakened compared to free passive scalar turbulence. For large Pr, \citet{Pandey:PF2016} and \citet{Pandey:PRE2016} have shown that $\epsilon_T \sim (U \Delta^2/d)\mathrm{Ra}^{-0.25}$ instead of $U \Delta^2/d$.}

To better understand the effects of walls, we need to study the behavior of thermal dissipation separately in the boundary layers and in the bulk. \textcolor{black}{It is generally believed that dissipation (thermal or viscous) occurs predominantly in the boundary layers~\cite{Baburaj:JFM2005,Baburaj:JFM2005a}. However, phenomenological arguments and numerical results presented by \citet{Verma:NJP2017} imply that significant dissipation occurs also at large scales, i.e., in the bulk. Motivated by this, Bhattacharya \textit{et al}.~\cite{Bhattacharya:PF2018} computed the viscous dissipation rate separately in the bulk and in the boundary layers for moderate Pr. Interestingly, they found the bulk dissipation to be greater, albeit marginally, than the boundary layer dissipation. On the other hand, the thermal dissipation for moderate Pr convection was shown to be dominant in the boundary layers; refer to Verzicco and Camussi~\cite{Verzicco:JFM2003} and \citet{Zhang:JFM2017}.}

\textcolor{black}{In this paper, we conduct a more detailed analysis of thermal dissipation rate in the bulk and boundary layers} for not only moderate Pr but also for large Pr convection.  Note that the statistics of thermal dissipation for large Pr are less explored in literature. We compare and quantify the total and average thermal dissipation rates in the bulk  and in the boundary layers using scaling arguments and numerical simulations. \textcolor{black}{We also} examine the probability distribution functions of the thermal dissipation in \textcolor{black}{these regions.} Our analysis is similar to that conducted by \citet{Bhattacharya:PF2018} on viscous dissipation rate.

The outline of the paper is as follows. In Sec.~\ref{sec:Equations}, we present the governing equations of RBC along with their nondimensionalization. We discuss the numerical method in Sec.~\ref{sec:Numerics}. In Sec.~\ref{sec:Numerical_Results}, we compute the thermal boundary layer thickness and present scaling arguments for the thermal dissipation rate in the bulk and in the boundary layers. We verify these scaling relations using our numerical results. We also study the spatial intermittency of thermal dissipation rate. Finally, we conclude in Sec.~\ref{sec:conclusion}. 

\section{Governing Equations} \label{sec:Equations}
In RBC, under the Boussinesq approximation, the thermal diffusivity ($\kappa$) and the kinematic viscosity ($\nu$) are treated as constants. The density of the fluid is considered to be a constant except for the buoyancy term in the governing equations. Further, the viscous dissipation term is considered to be small and is therefore dropped from the temperature equation. 
The governing equations of RBC are as follows\cite{Chandrasekhar:book:Instability,Verma:book:BDF}:
\bea
\frac{\partial \mathbf{u}}{\partial t} + (\mathbf{u} \cdot \nabla) \mathbf{u} & = & -\nabla p/\rho_0 + \alpha g T \hat{z} + \nu \nabla^2 \mathbf{u}, \label{eq:momentum}\\
\frac{\partial T}{\partial t} + (\mathbf{u} \cdot \nabla) T & = & \kappa \nabla^2 T, \label{eq:energy}  \\
\nabla \cdot \mathbf{u} &=& 0, \label{eq:continuity} 
\eea  
where $\mathbf{u}$ and $p$ are the velocity and pressure fields respectively, $T$ is the temperature field with respect to a reference temperature, $\alpha$ is the thermal expansion coefficient, $\rho_0$ is the mean density of the fluid, and $g$ is acceleration due to gravity. 

Using $d$ as the length scale, $\sqrt{\alpha g \Delta d}$ as the velocity scale, and $\Delta$ as the temperature scale, we non-dimensionalize Eqs.~(\ref{eq:momentum})-(\ref{eq:continuity}), which yields
\bea
\frac{\partial \mathbf{u}}{\partial t} + \mathbf{u} \cdot \nabla \mathbf{u} &=& -\nabla p + T \hat{z} +  \mathrm{\sqrt{\frac{Ra}{Pr}}}\nabla^2 \mathbf{u}, \label{eq:NDMomentum} \\
\frac{\partial T}{\partial t} + \mathbf{u}\cdot \nabla T &=& \frac{1}{\sqrt{\mathrm{Ra Pr}}}\nabla^2 T. \label{eq:NDTheta}\\
\nabla \cdot \mathbf{u} &=& 0, \label{eq:NDContinuity}
\eea 

In Sec.~\ref{sec:Numerics}, we describe the numerical method used for our simulations.

\section{Numerical Method} \label{sec:Numerics}
We conduct our numerical analysis for (i) $\mathrm{Pr=1}$ and (ii) $\mathrm{Pr=100}$ fluids. For $\mathrm{Pr=1}$, we use the simulation data of \citet{Bhattacharya:PF2018} and \citet{Kumar:RSOS2018}, which were obtained using the finite volume code OpenFOAM~\cite{Jasak:CD2007}. The simulations were conducted on a $256^3$ grid for \textcolor{black}{Ra} ranging from $10^6$ to $10^8$. No-slip boundary conditions were imposed at all the walls, isothermal boundary conditions at the top and bottom walls, and adiabatic boundary conditions at the sidewalls. For time marching, \textcolor{black}{second-order} Crank-Nicholson scheme was used. For $\mathrm{Pr=100}$, we conduct fresh simulations following the aforementioned schemes, boundary conditions, and grid resolution for Ra's ranging from $\mathrm{2 \times 10^6}$ to $\mathrm{5 \times 10^7}$. \textcolor{black}{A constant time-step was chosen, with $\Delta t=10^{-3}$ and $5 \times 10^{-4}$, depending on the parameters~(see Table~\ref{table:SimDetails} for details). Here, $t=1$ corresponds to $d/\sqrt{\alpha g \Delta d}$.}

We ensure that a minimum of 8 grid points is in the thermal boundary layers, thereby satisfying the resolution criterion set by \citet{Grotzbach:JCP1983} and \citet{Verzicco:JFM2003}. \textcolor{black}{In RBC, the thermal boundary layer thickness $\delta_T$ is defined as the distance between the wall and the point where the tangent to the planar-averaged temperature profile near the wall intersects with the average bulk temperature line~\cite{Ahlers:RMP2009,Scheel:JFM2012, Shi:JFM2012, Scheel:JFM2014}. To ensure that the smallest length scales are resolved, we note} that the ratio of the Batchelor length scale~\cite{Batchelor:JFM1959_largeSc} $\eta_\theta=(\nu \kappa^2/\epsilon_u)^{1/4}$ to the maximum mesh width $\Delta x_{\mathrm{max}}$ remains greater than unity for all runs. The only exception is for $\mathrm{Ra}=10^8$, $\mathrm{Pr}=1$ case where $\eta_\theta = 0.8$, which is marginally less than unity. \textcolor{black}{The Nusselt numbers computed using our data are consistent with those obtained in other simulations of RBC for the same geometry~\cite{Wagner:PF2013,Pandey:PF2016,Pandey:PRE2016}; this is how we validate our data.} Further, the \textcolor{black}{Nusselt numbers} computed numerically using $\langle u_z T \rangle$ match closely with those computed using $\epsilon_T$ and  Eq. (\ref{eq:SS_exact}). This is further validates our simulations. See Table~\ref{table:SimDetails} for the comparison of these two Nusselt numbers. All the quantities analyzed in this work are time-averaged over 40-100 snapshots after attaining steady-state~(see Table~\ref{table:SimDetails}). 
\begin{table*}
\caption{Details of our \textcolor{black}{numerical data obtained using} direct numerical simulations performed in a cubical box: the Prandtl number (Pr), the Rayleigh Number (Ra),  the P\'{e}clet Number (Pe), \textcolor{black}{the time-step ($\Delta t$),} the ratio of the Batchelor length scale~\cite{Batchelor:JFM1959_largeSc} ($\eta_\theta$) to the maximum mesh width $\Delta x_{\mathrm{max}}$, the Nusselt Number (Nu), the Nusselt number ($\mathrm{Nu_S}$) deduced from $\epsilon_T$ using Eq.~(\ref{eq:SS_exact}), the ratio of the thermal boundary layer thickness $\delta_T$ to the cell height $d$, the number of grid points in the thermal boundary layer ($N_{\mathrm{BL}}$), and the number of snapshots over which the quantities are averaged.}
\begin{ruledtabular}
\begin{tabular}{c c c c c c c c c c}
$\mathrm{Pr}$ & $\mathrm{Ra}$ & $\mathrm{Pe}$ & \textcolor{black}{$\Delta t$} & $\eta_\theta/\Delta x_{\mathrm{max}}$ & $\mathrm{Nu}$ & $\mathrm{Nu_S}$ & $\delta_T/d$ & $N_{\mathrm{BL}}$ & Snapshots\\
\hline
$1$ & $\mathrm{1 \times 10^6}$ & $\mathrm{150}$ & \textcolor{black}{$1 \times 10^{-3}$} & $3.6$ & $8.40$ & $8.26$ & $0.061$ & $23$ & $56$\\
$1$ & $\mathrm{2 \times 10^6}$ & $\mathrm{212}$ & \textcolor{black}{$1 \times 10^{-3}$} & $2.8$ & $10.1$ & $10.1$ & $0.050$ & $19$ & $56$\\
$1$ & $\mathrm{5 \times 10^6}$ & $\mathrm{342}$ & \textcolor{black}{$1 \times 10^{-3}$} & $2.1$ & $13.3$ & $13.4$ & $0.037$ & $14$ & $55$\\
$1$ & $\mathrm{1 \times 10^7}$ & $\mathrm{460}$ & \textcolor{black}{$1 \times 10^{-3}$} & $1.7$ & $16.0$ & $16.1$ & $0.031$ & $12$ & $100$\\
$1$ & $\mathrm{2 \times 10^7}$ & $\mathrm{654}$ & \textcolor{black}{$1 \times 10^{-3}$} & $1.3$ & $20.0$ & $19.7$ & $0.025$ & $10$ & $100$\\
$1$ & $\mathrm{5 \times 10^7}$ & $\mathrm{1080}$ & \textcolor{black}{$1 \times 10^{-3}$} & $1.0$ & $25.5$ & $25.7$ & $0.019$ & $8$ & $101$\\
$1$ & $\mathrm{1 \times 10^8}$ & $\mathrm{1540}$ & \textcolor{black}{$1 \times 10^{-3}$} & $0.8$ & $32.8$ & $32.0$ & $0.016$ & $7$ & $86$\\
$100$ & $\mathrm{2 \times 10^6}$ & $\mathrm{277}$ & \textcolor{black}{$5 \times 10^{-4}$} & $2.8$ & $11.1$ & $11.1$ & $0.045$ & $17$ & $41$\\
$100$ & $\mathrm{5 \times 10^6}$ & $\mathrm{496}$ & \textcolor{black}{$1 \times 10^{-3}$} & $2.0$ & $14.5$ & $14.4$ & $0.034$ & $13$ & $50$\\
$100$ & $\mathrm{1 \times 10^7}$ & $\mathrm{698}$ & \textcolor{black}{$1 \times 10^{-3}$} & $1.6$ & $17.2$ & $17.1$ & $0.029$ & $12$ & $52$\\
$100$ & $\mathrm{2 \times 10^7}$ & $\mathrm{1036}$ & \textcolor{black}{$1 \times 10^{-3}$} & $1.3$ & $20.1$ & $20.3$ & $0.025$ & $10$ & $99$\\
$100$ & $\mathrm{5 \times 10^7}$ & $\mathrm{1772}$ & \textcolor{black}{$1 \times 10^{-3}$} & $1.0$ & $26.0$ & $26.0$ & $0.019$ & $8$ & $101$
\end{tabular}
\label{table:SimDetails}
\end{ruledtabular}
\end{table*}

In the Sec.~\ref{sec:Numerical_Results}, \textcolor{black}{we discuss} the numerical results, focussing on the scaling of the thermal dissipation rate in the bulk and in the boundary layers, their relative contributions to the total thermal dissipation rate, and their spatial intermittency. 

\section{Numerical Results} \label{sec:Numerical_Results}

\subsection{Boundary layer thickness} \label{subsec:BL_thickness}
\begin{figure}[b]
\includegraphics[scale=0.45]{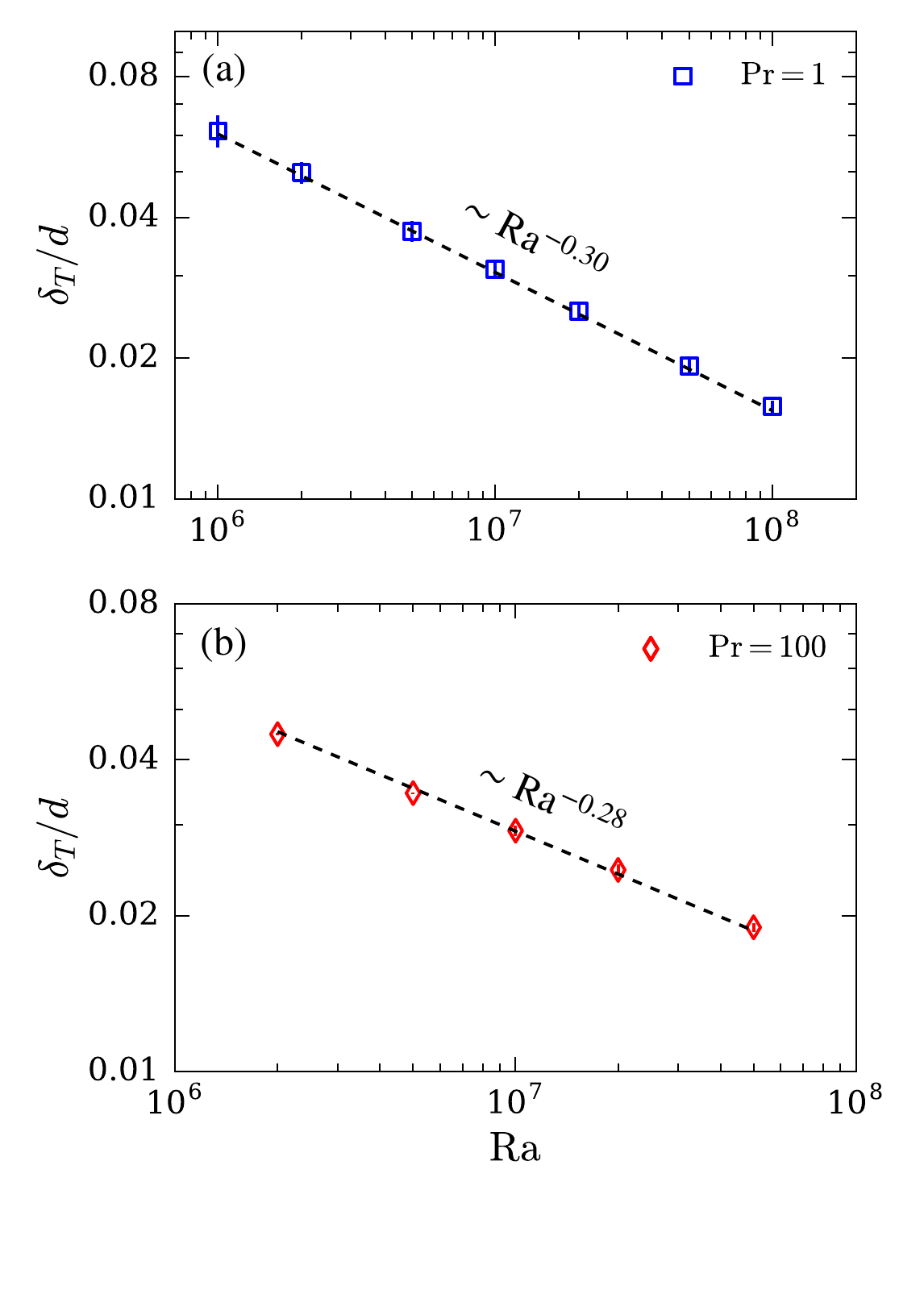}
\caption{\textcolor{black}{For (a) $\mathrm{Pr=1}$ and (b) $\mathrm{Pr=100}$: plots of normalized thermal boundary layer thickness $\delta_T/d$ vs. $\mathrm{Ra}$, along with $\mathrm{Ra}^{-0.30}$ and $\mathrm{Ra}^{-0.28}$ fits (dashed curves). The error-bars represent the standard deviation of the dataset with respect to the temporal average.}}
\label{fig:BL_thickness}
\end{figure} 
\textcolor{black}{Using the simulation results, we first compute the thickness of the thermal boundary layers.} Theoretically, boundary layer thickness ($\delta_T$) is related to the Nusselt number as~\cite{Ahlers:RMP2009} 
\be
\frac{\delta_T}{d} = \frac{1}{2\mathrm{Nu}}.
\label{eq:delta_T_PB_0}
\ee
Now, as discussed in Sec.~\ref{sec:Introduction}, $\mathrm{Nu} \sim \mathrm{Ra}^{0.3}$ for $\mathrm{Pr}$ \textcolor{black}{of order} 1. Numerical simulations~\cite{Pandey:PRE2014,Pandey:PF2016,Pandey:PRE2016} reveal that $\mathrm{Nu} \sim \mathrm{Ra}^{0.3}$ for large Pr as well. Therefore, for both $\mathrm{Pr=1}$ and $100$, we expect
\be
\frac{\delta_T}{d} \sim \mathrm{Ra}^{-0.3}.
\label{eq:delta_T_PB}
\ee
We numerically compute $\delta_T$'s using the planar averaged temperature profile and list them in Table~\ref{table:SimDetails}. Further, we plot them  versus Ra in \textcolor{black}{Figs.~\ref{fig:BL_thickness}(a) and (b)} for both  $\mathrm{Pr}=1$ and $100$. The best-fit curves of the data yield \textcolor{black}{
\begin{equation}
\frac{\delta_T}{d} =
\begin{cases}
3.6 \mathrm{Ra}^{-0.30}, \quad \mathrm{Pr=1}, \\
2.4 \mathrm{Ra}^{-0.28}, \quad \mathrm{Pr=100},
\end{cases}
\label{eq:delta_T_Sim}
\end{equation} }
with the error in the exponents being approximately $0.01$. The obtained fit is \textcolor{black}{reasonably} consistent with Eq.~(\ref{eq:delta_T_PB}).

\subsection{Scaling of thermal dissipation rate}
In this subsection, \textcolor{black}{we study} the scaling of average thermal dissipation rate in the bulk  ($\epsilon_{T, \mathrm{bulk}}$) and in the boundary layers ($\epsilon_{T,\mathrm{BL}}$) using our numerical data. These quantities are dissipation per unit volume. Based on these, using scaling arguments, \textcolor{black}{we predict} the relations for the total dissipation rate in the bulk ($\tilde{D}_{T,\mathrm{bulk}}$) and in the boundary layers ($\tilde{D}_{T,\mathrm{BL}}$), which are the products of average thermal dissipation rates in these regions and their corresponding volumes. \textcolor{black}{We verify} their scaling relations using our simulation data and analyze the relative strength of the bulk and the boundary layer dissipation.
\subsubsection{Bulk dissipation}
\begin{figure}[htbp]
\includegraphics[scale=0.45]{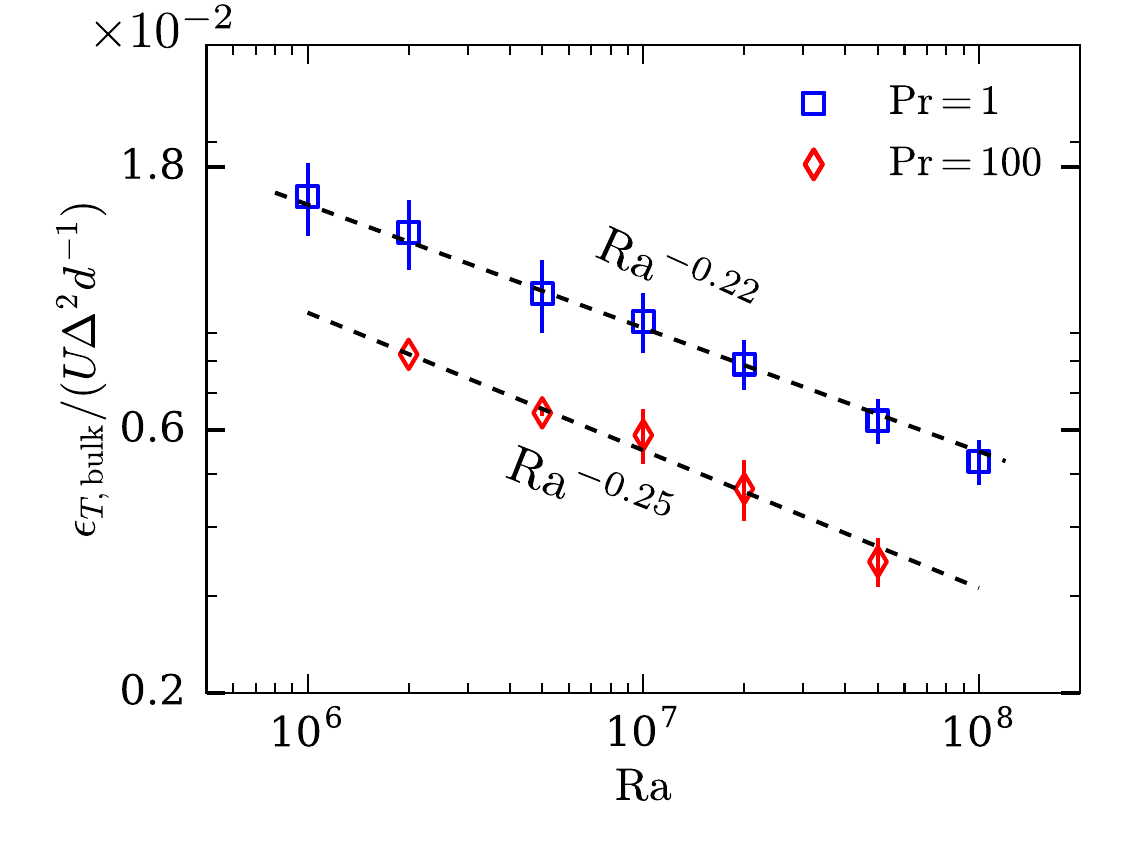}
\caption{Plots of average thermal dissipation rate in the bulk, normalized with $U \Delta^2 /d$, versus Ra. The bulk dissipation is distinctly weaker than $U \Delta^2/d$. The error bars represent the standard deviation of the dataset with respect to the temporal average.}
\label{fig:AverageDissipation}
\end{figure}

\textcolor{black}{Using our simulation data, we  numerically compute $\epsilon_{T, \mathrm{bulk}} = \langle \kappa |\nabla T(\mathbf{r})|^2 \rangle_\mathrm{bulk}$ and the large-scale mean flow $U=\sqrt{\langle |\mathbf{u} (\mathbf{r})|^2 \rangle }$.}
In deriving their unifying scaling theory, Grossmann and Lohse~\cite{Grossmann:JFM2000,Grossmann:PRL2001} argued that $\epsilon_{T,\mathrm{bulk}} \sim U \Delta^2/d$. However, from our numerical data, we observe that
\be
\epsilon_{T,\mathrm{bulk}} \sim
\begin{cases}
(U \Delta^2/d) \mathrm{Ra}^{-0.22}, \quad \mathrm{Pr=1}, \\
(U \Delta^2/d) \mathrm{Ra}^{-0.25}, \quad \mathrm{Pr=100},
\end{cases}
\label{eq:epsilon_T_bulk}
\ee
instead of $U\Delta^2/d$~(see Fig.~\ref{fig:AverageDissipation}). The errors in the
exponents are 0.02 and 0.01 for $\mathrm{Pr=1}$ and $100$ respectively. \textcolor{black}{Thus, the
thermal dissipation in the bulk in RBC scales similar to the dissipation in the entire volume, and is distinctly weaker than that in passive
scalar turbulence.} For moderate Pr fluids, the decrease of
$\epsilon_{T,\mathrm{bulk}}/(U\Delta^2/d)$ with Ra has also been observed by
\citet{Emran:JFM2008} and \citet{Verzicco:JFM2003} for convection in a cylindrical
cell, and by \citet{Zhang:JFM2017} for two-dimensional RBC. 
\textcolor{black}{As discussed in Sec.~\ref{sec:Introduction}, the walls suppress nonlinear interactions in RBC~\cite{Pandey:PF2016,Pandey:PRE2016}, consequently weakening the thermal dissipation rate at large scales.} Note that
\citet{Bhattacharya:PF2018} observed similar suppression of viscous dissipation in
the bulk, where $\epsilon_{u,\mathrm{bulk}} \sim (U^3/d) \mathrm{Ra}^{-0.18}$
instead of $U^3/d$ for $\mathrm{Pr=1}$. 

The aforementioned suppression has an important implication in the scaling of the total thermal dissipation in the bulk ($\tilde{D}_{T,\mathrm{bulk}}$). The bulk volume can be approximated as
\begin{equation}
V_\mathrm{bulk} = (d-2\delta_T)d^2 \approx d^3,
\label{eq:BulkVolume}
\end{equation}
because $\delta_T \ll d$ (see Table~\ref{table:SimDetails}). We will now derive the scaling relations for $\tilde{D}_{T,\mathrm{bulk}}$ separately for $\mathrm{Pr}=1$ and $100$.

\begin{enumerate}

\item $\mathrm{Pr=1}$: Using Eqs.~(\ref{eq:epsilon_T_bulk}) and (\ref{eq:BulkVolume}), we write the following for the bulk dissipation:
\bea
\tilde{D}_{T,\mathrm{bulk}} = \epsilon_{T,\mathrm{bulk}}V_\mathrm{bulk} 
\sim \left(\frac{U \Delta^2}{d} \mathrm{Ra}^{-0.22} \right) d^3. 
\label{eq:D_T_bulk_P1_1} 
\eea
By multiplying the numerator and the denominator of the rightmost expression in Eq.~(\ref{eq:D_T_bulk_P1_1}) by $d/\kappa$, we rewrite $\tilde{D}_{T,\mathrm{bulk}}$ as
\bea
\left(\frac{U \Delta^2}{d} \mathrm{Ra}^{-0.22} \right) d^3 = (\kappa \Delta^2 d)\mathrm{PeRa^{-0.22}},
\label{eq:D_T_bulk_P1_2}
\eea
where $\mathrm{Pe}=Ud/\kappa$ is the P\'{e}clet number. As discussed in Sec.~\ref{sec:Introduction}, $\mathrm{Pe} \sim \mathrm{Ra}^{0.5}$ for moderate Pr. Substituting this relation in Eqs.~(\ref{eq:D_T_bulk_P1_1})  and (\ref{eq:D_T_bulk_P1_2}), we obtain 
\begin{equation}
\tilde{D}_{T,\mathrm{bulk}} \sim  \textcolor{black}{(\kappa \Delta^2 d)}\mathrm{Ra}^{0.28}.
\label{eq:D_T_bulk_prediction}
\end{equation}

\item $\mathrm{Pr=100}$: Applying a similar procedure, we can write the total dissipation in the bulk for $\mathrm{Pr=100}$ as
\begin{equation}
\tilde{D}_{T,\mathrm{bulk}} \sim (\kappa \Delta^2 d) \mathrm{PeRa^{-0.25}},
\label{eq:D_T_bulk_P100_1}
\end{equation}
because $\epsilon_{T,\mathrm{bulk}} \sim (U\Delta^2/d)\mathrm{Ra}^{-0.25}$ in this case. Now, according to the predictions of \citet{Grossmann:PRL2001} and \citet{Shishkina:PRF2017} for large Pr convection, $\mathrm{Pe} \sim \mathrm{Ra}^{3/5}$. \citet{Pandey:PRE2014}, \citet{Pandey:PF2016}, and \citet{Pandey:PRE2016} have also shown that for large Pr, $\mathrm{Pe \sim Ra}^{0.6}$. Substituting this relation in Eq.~(\ref{eq:D_T_bulk_P100_1}), we obtain
\begin{equation}
\tilde{D}_{T,\mathrm{bulk}} \sim \textcolor{black}{(\kappa \Delta^2 d)} \mathrm{Ra}^{0.35}.
\label{eq:D_T_bulk_prediction_P100}
\end{equation}
\end{enumerate}

Thus, the suppression of thermal dissipation in the bulk leads to a weaker \textcolor{black}{dependence} of the total thermal dissipation with Ra. Note that in the absence of this suppression, $\tilde{D}_{T,\mathrm{bulk}} \sim \textcolor{black}{(\kappa \Delta^2 d)}\mathrm{Pe}$. Had this been the case, $\tilde{D}_{T,\mathrm{bulk}}$, \textcolor{black}{normalized with $\kappa \Delta^2 d$, would have been proportional to $\mathrm{Ra}^{0.5}$ for $\mathrm{Pr}=1$ and $\mathrm{Ra}^{0.6}$ for $\mathrm{Pr}=100$.}
\subsubsection{Boundary layer dissipation}
\begin{figure}[htbp]
\includegraphics[scale=0.45]{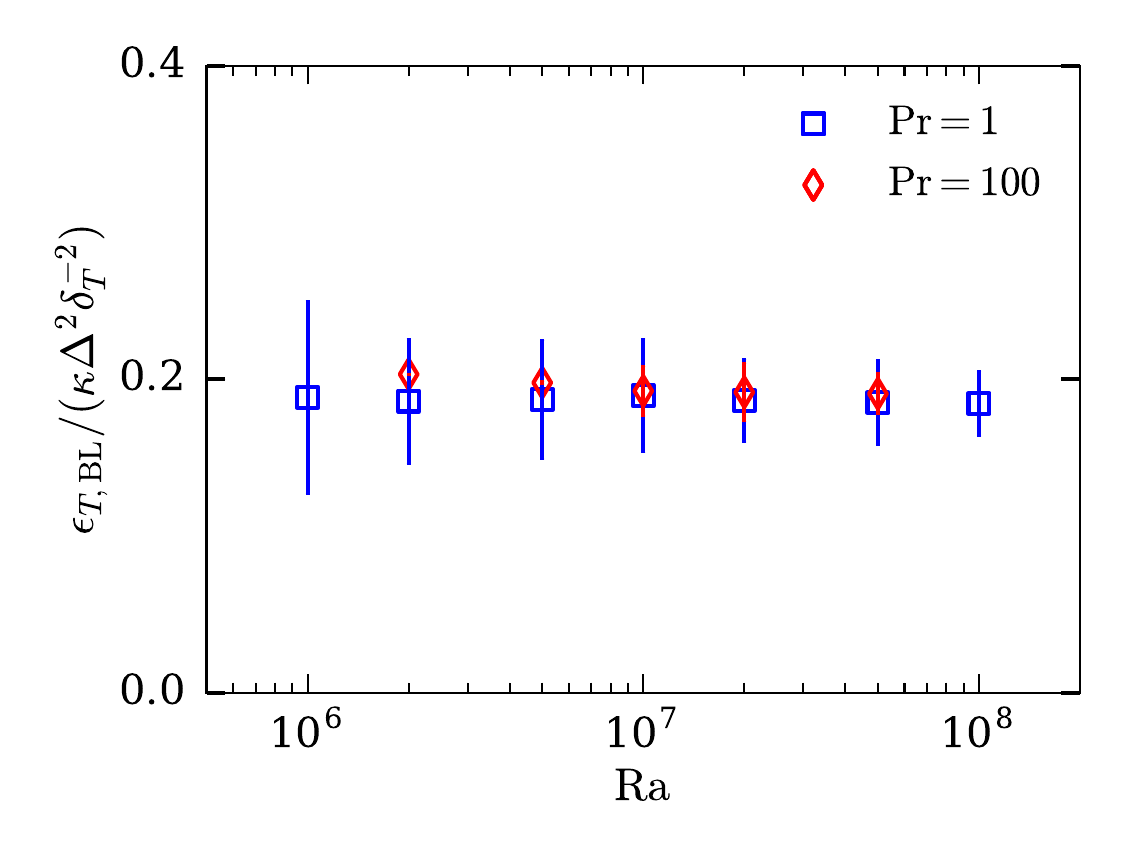}
\caption{Plots of average thermal dissipation in the boundary layers, normalized with $\kappa \Delta^2 / \delta_T^2$, versus Ra. The error bars represent the standard deviation of the dataset with respect to the temporal average.}
\label{fig:AverageDissipation_BL}
\end{figure}
The heat transport in the boundary layers is primarily diffusive due to steep temperature gradients. Thus, we expect the thermal dissipation in the boundary layers to \textcolor{black}{be given by}
\begin{equation}
\epsilon_{T,\mathrm{BL}} \sim \kappa \Delta^2/\delta_T^2.
\label{eq:epsilon_T_BL}
\end{equation}
We verify this by plotting the numerically computed $\epsilon_{T,\mathrm{BL}}/(\kappa \Delta^2/ \delta_T^{2})$ versus $\mathrm{Ra}$ in Fig.~\ref{fig:AverageDissipation_BL}, where we observe the \textcolor{black}{slope} to be flat. For $\mathrm{Pr=100}$ and at lower \textcolor{black}{Ra}, however, there is a very slight decrease of $\epsilon_{T,\mathrm{BL}}/(\kappa \Delta^2/ \delta_T^{2})$ with Ra. However, we will ignore this in our scaling analysis. 

The total thermal dissipation in the boundary layers is given by $\tilde{D}_{T,\mathrm{BL}}=\epsilon_{T,\mathrm{BL}}V_\mathrm{BL}$. Substituting Eq.~(\ref{eq:epsilon_T_BL}) in the above relation and noting that $V_\mathrm{BL} = 2\delta_T d^2$, we obtain
\begin{equation}
\tilde{D}_{T,\mathrm{BL}} \sim \left(\frac{\kappa \Delta^2}{\delta_T^2} \right) \delta_T d^2 \sim \kappa \Delta^2 d \left(\frac{d}{\delta_T} \right).
\label{eq:D_T_BL_1}
\end{equation}
As discussed in Sec.~\ref{subsec:BL_thickness}, \textcolor{black}{$\delta_T/d \sim \mathrm{Ra}^{-0.30}$ for $\mathrm{Pr}=1$ and $\sim \mathrm{Ra}^{-0.28}$ for $\mathrm{Pr}=100$}. Substituting these relations in Eq.~(\ref{eq:D_T_BL_1}), we obtain 
\begin{equation}
\textcolor{black}{\tilde{D}_{T,\mathrm{BL}} \sim
\begin{cases} 
(\kappa \Delta^2 d)\mathrm{Ra}^{0.30}, \quad \mathrm{Pr=1} \\
(\kappa \Delta^2 d)\mathrm{Ra}^{0.28}, \quad \mathrm{Pr=100}
\end{cases}}
\label{eq:D_T_BL_prediction}
\end{equation}

\subsubsection{Ratio of the boundary layer and the bulk dissipation}
To analyze the relative strengths of the thermal dissipation in the bulk and in the boundary layers, we divide Eq.~(\ref{eq:D_T_BL_prediction}) with Eqs.~(\ref{eq:D_T_bulk_prediction}) and (\ref{eq:D_T_bulk_prediction_P100}) to obtain the ratio of the total dissipation in the boundary layers and the bulk for $\mathrm{Pr}=1$ and $100$ respectively. The {\color{black} predicted} ratio is 
\be
\textcolor{black}{
\frac{\tilde{D}_{T,\mathrm{BL}}}{\tilde{D}_{T,\mathrm{bulk}}} \sim
\begin{cases}
 \mathrm{Ra}^{0.02}, \quad \mathrm{Pr=1}, \\
 \mathrm{Ra}^{-0.07}, \quad \mathrm{Pr=100}.
\end{cases}}
\label{eq:D_T_ratio}
\ee

\textcolor{black}{Thus, we expect the ratio of the boundary layer and bulk dissipation to have a  weak dependence on Ra}. For $\mathrm{Pr=1}$, this ratio remains approximately constant, implying that the relative strengths of the bulk and the boundary layer dissipation remain roughly invariant with Ra. However, for $\mathrm{Pr=100}$, the above ratio decreases weakly with Ra; this implies that the relative strength of the boundary layer dissipation decreases with Ra and that of the bulk dissipation increases with Ra. The magnitudes of the prefactors in Eq.~(\ref{eq:D_T_ratio}) determine whether the bulk or the boundary layer dissipation is dominant. These prefactors are obtained using numerical simulations.

\subsubsection{Numerical verification of the scaling arguments}
\begin{figure*}[htbp]
\includegraphics[scale=0.42]{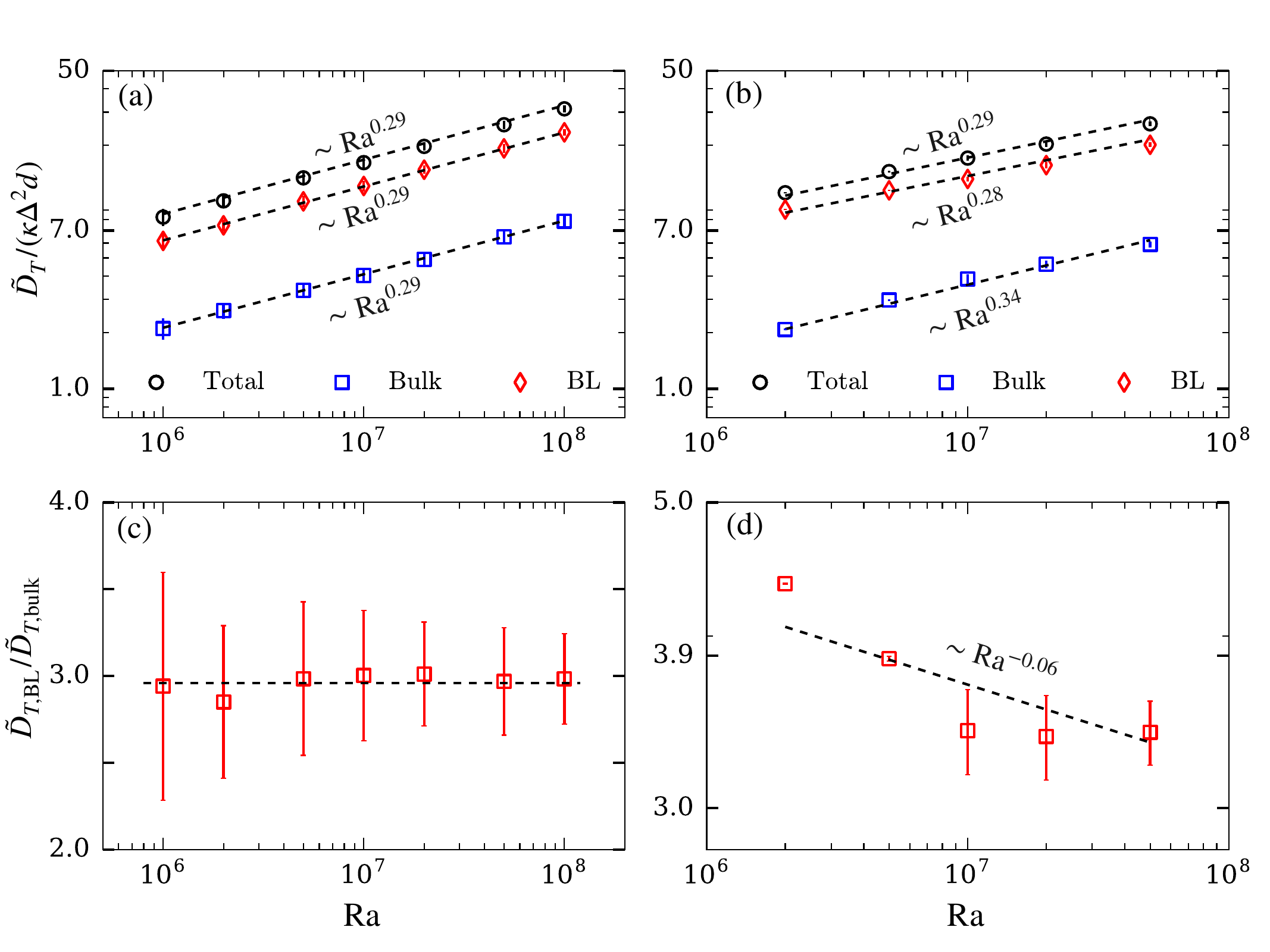}
\caption{For (a) $\mathrm{Pr=1}$ and (b) $\mathrm{Pr=100}$: plots of thermal dissipation rates $\tilde{D}_T$---total,  bulk, and in the boundary layers~(BL)---vs.~$\mathrm{Ra}$. For (c) $\mathrm{Pr=1}$ and (d) $\mathrm{Pr=100}$: plots of the dissipation rate ratio, $\tilde{D}_{T,\mathrm{BL}}/\tilde{D}_{T, \mathrm{bulk}}$, vs.~Ra. The error bars represent the standard deviation of the dataset with respect to the temporal average.}
\label{fig:contribution}
\end{figure*}

We numerically verify the scaling relations predicted by Eqs.~(\ref{eq:D_T_bulk_prediction}), (\ref{eq:D_T_bulk_prediction_P100}), (\ref{eq:D_T_BL_prediction}), and (\ref{eq:D_T_ratio}). We compute $\tilde{D}_T$ (the total dissipation in the entire volume), $\tilde{D}_{T,\mathrm{bulk}}$, and $\tilde{D}_{T,\mathrm{BL}}$ using our simulation data and plot them versus Ra in Fig.~\ref{fig:contribution}(a) for $\mathrm{Pr=1}$ and in Fig.~\ref{fig:contribution}(b) for $\mathrm{Pr=100}$. Our data fits well with following \textcolor{black}{expressions}:
\begin{align}
\tilde{D}_{T} &=0.16 \textcolor{black}{c}\mathrm{Ra}^{0.29}, \quad \mathrm{Pr=1,100}, 
\label{eq:DT_Scaling} \\
\tilde{D}_{T,\mathrm{bulk}} &=
\begin{cases}
0.041 \textcolor{black}{c}\mathrm{Ra}^{0.29}, \quad \mathrm{Pr=1}, \\
0.015 \textcolor{black}{c}\mathrm{Ra}^{0.34}, \quad \mathrm{Pr=100},
\end{cases} 
\label{eq:DT_Bulk_Scaling} \\
\tilde{D}_{T,\mathrm{BL}} &=
\begin{cases}
0.12 \textcolor{black}{c}\mathrm{Ra}^{0.29}, \quad \mathrm{Pr=1}, \\
0.15 \textcolor{black}{c}\mathrm{Ra}^{0.28}, \quad \mathrm{Pr=100},
\end{cases} 
\label{eq:DT_BL_Scaling}
\end{align}
\textcolor{black}{where $c=\kappa \Delta^2 d$. The errors in the exponents in the above expressions} range from 0.001 to 0.02. The above expressions match with the scaling arguments presented in Eqs.~(\ref{eq:D_T_BL_prediction}), (\ref{eq:D_T_bulk_prediction}), and (\ref{eq:D_T_bulk_prediction_P100}) within the fitting error. 

The computed ratio of the boundary layer and the bulk dissipation is
\be
\frac{\tilde{D}_{T,\mathrm{BL}}}{\tilde{D}_{T,\mathrm{bulk}}}  \approx
\begin{cases}
3.0,& \mathrm{Pr=1}, \\
10 \mathrm{Ra}^{-0.06},& \mathrm{Pr=100}, 
\end{cases}
\label{eq:DT_BL_Bk_Ratio_sim}
\ee
which agrees well with Eq.~(\ref{eq:D_T_ratio}). We plot this ratio in Fig.~\ref{fig:contribution}(c) for $\mathrm{Pr=1}$ and Fig.~\ref{fig:contribution}(d) for $\mathrm{Pr=100}$. 

\begin{figure}[t]
\includegraphics[scale=0.2]{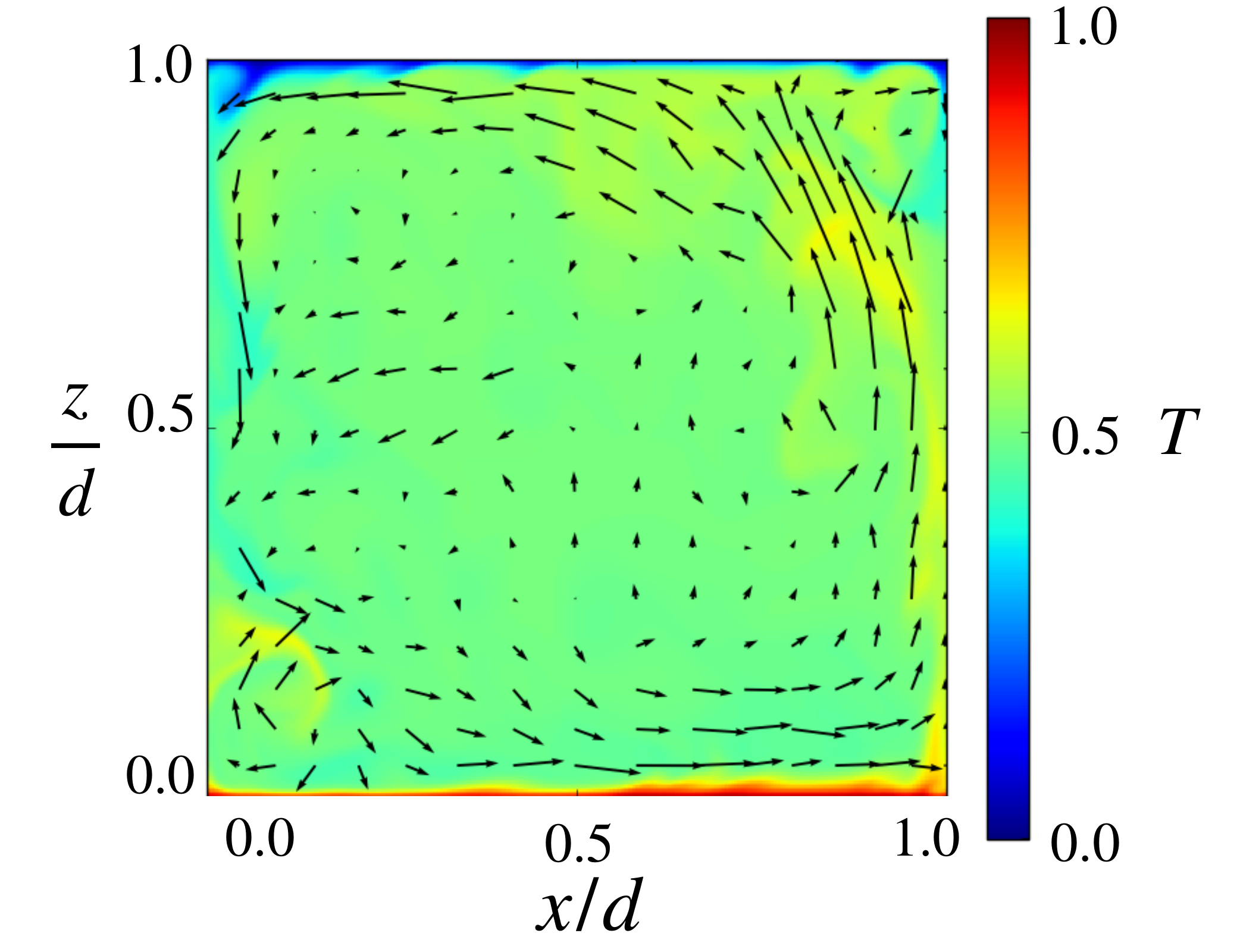}
\caption{\textcolor{black}{For $\mathrm{Ra=10^8}$, $\mathrm{Pr=1}$: Density plot of temperature field, superimposed with velocity vector plot, on $x$-$z$ plane at $y=d/2$. The temperature is approximately constant in the bulk, unlike the velocity fluctuations.}}
\label{fig:density}
\end{figure}

Because of the prefactors in Eq.~(\ref{eq:DT_BL_Bk_Ratio_sim}), the ratio of the boundary layer and the bulk dissipation remains above unity, implying that the boundary layer dissipation is larger than the bulk dissipation, \textcolor{black}{although they are of the same order}. As shown in Figs.~\ref{fig:contribution}(c) and \ref{fig:contribution}(d), the boundary layer dissipation is approximately 3-4 times greater than the bulk dissipation. This is unlike viscous dissipation \textcolor{black}{for $\mathrm{Pr=1}$,} where the dissipation in the bulk is greater, albeit marginally, than that in the boundary layers~\cite{Bhattacharya:PF2018}. \textcolor{black}{This is because while the temperature is fairly constant in the bulk (except for a few regions of localized plumes), the velocity in the bulk is not so, as illustrated in Fig.~\ref{fig:density}. Here, we show the temperature density plot superimposed with velocity vector plot on $x$-$z$ plane at $y=d/2$, for $\mathrm{Ra=10^8}$, $\mathrm{Pr=1}$. Clearly, the velocity fluctuations are large near the walls (just outside the viscous boundary layers) but small near the center. On the other hand, $T \approx 0.5$ in the bulk. Thus, the velocity gradients in the bulk are more pronounced than the temperature gradients; this results in stronger viscous dissipation compared to thermal dissipation in the bulk. However, one must note that for $\mathrm{Pr=100}$, the viscous boundary layers will occupy almost the entire volume; thus the viscous dissipation in the boundary layers will be dominant. Also, we need to carefully simulate low Pr convection to find out whether bulk or boundary layer dissipation dominates in this regime.}

The dominance of the total thermal dissipation in the boundary layers has been reported previously for convection in a slender cylindrical cell~\cite{Verzicco:JFM2003} and for two-dimensional convection~\cite{Zhang:JFM2017}.

\subsection{Spatial intermittency of thermal dissipation rate}
\begin{figure}[t]
\includegraphics[scale=0.45]{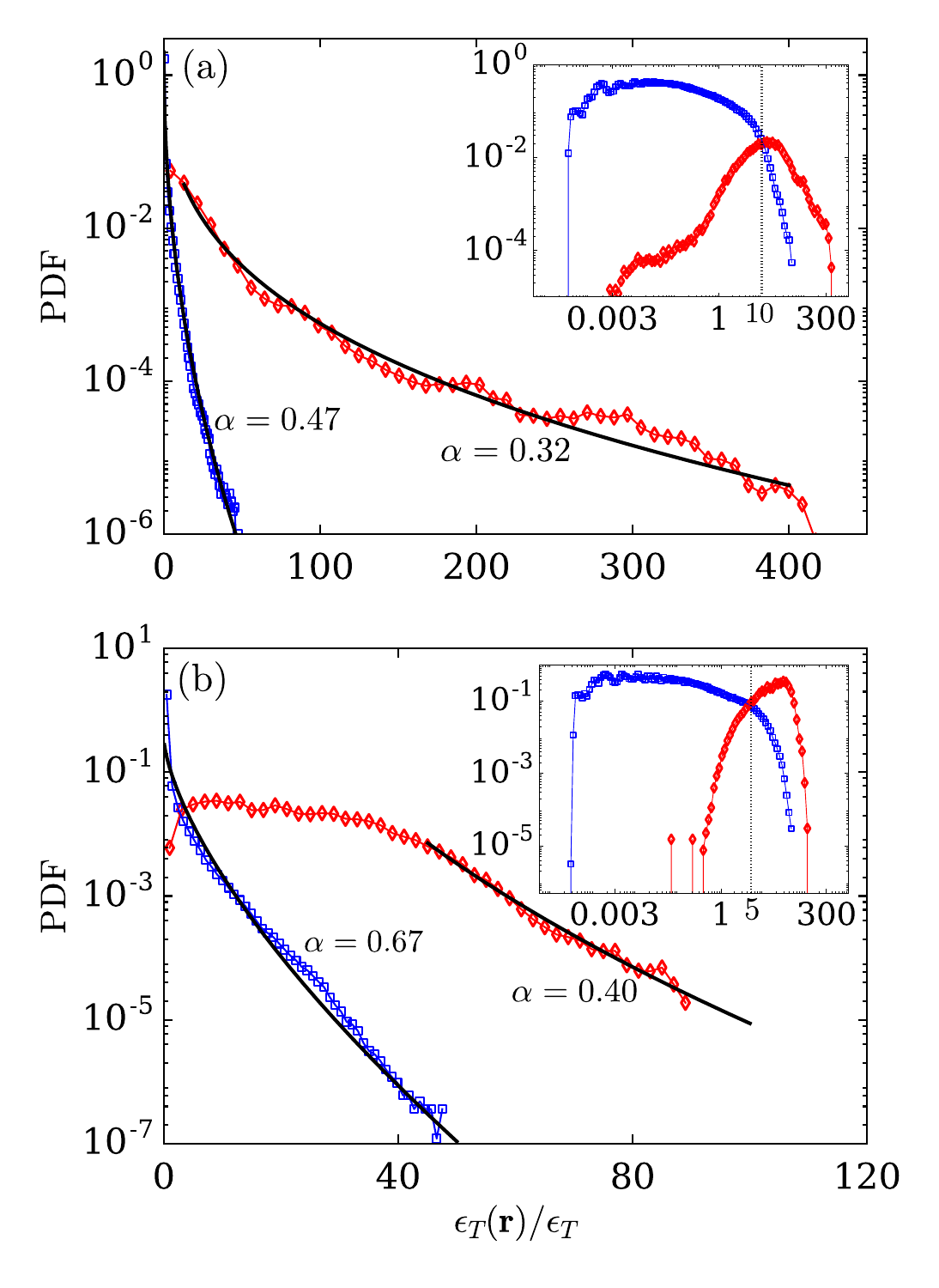}
\caption{For $\mathrm{Ra}=5 \times 10^7$ and (a) $\mathrm{Pr}=1$ and (b) $\mathrm{Pr=100}$: probability distribution functions (PDF) of normalized local dissipation rate $\epsilon_T(\mathbf{r})/\epsilon_T$  in the bulk (black squares) and in the boundary layers (red diamonds). The PDFs of both $\epsilon_{T,\mathrm{BL}}$ and $\epsilon_{T,\mathrm{bulk}}$ fit well with stretched exponential curves (black solid lines). The insets in (a) and (b) show the log-log plots of the PDFs of $\epsilon_T(\mathbf{r})$ in the bulk and the boundary layers.}
\label{fig:PDF}
\end{figure}
In this subsection, we will study the intermittency of the local thermal dissipation rate $\epsilon_T(\mathbf{r})$. \textcolor{black}{Since $\delta_T / d \ll 1$~(see Fig.~\ref{fig:BL_thickness}), the boundary layers occupy a much smaller volume than the bulk. Therefore, $\epsilon_T({\bf r})$ is much stronger in the boundary layers than in the bulk.} 
  
We compute the probability distribution functions (PDF) of $\epsilon_T^*(\mathbf{r})=\epsilon_T(\mathbf{r})/\epsilon_T$ in the entire volume, bulk and boundary layers to quantify the spatial intermittency of thermal dissipation rate. The PDFs are computed for $\mathrm{Ra}=5 \times 10^7$ for both $\mathrm{Pr=1}$ and 100. We plot these quantities in Fig.~\ref{fig:PDF}(a) for $\mathrm{Pr}=1$ and in Fig.~\ref{fig:PDF}(b) for $\mathrm{Pr}=100$. From the inset of Fig.~\ref{fig:PDF}(a), we observe that for $\mathrm{Pr=1}$, $P(\epsilon^*_{T,\mathrm{bulk}}) \gg P(\epsilon^*_{T,\mathrm{BL}})$ for $\epsilon_T^* < 10$, while $P(\epsilon^*_{T,\mathrm{bulk}}) \ll P(\epsilon^*_{T,\mathrm{BL}})$ for $\epsilon_T^* > 10$. This clearly shows that thermal dissipation is weak in the bulk and strong in the boundary layers. We observe a similar behaviour for Pr=100, but with cut-off $\epsilon_T^* \approx 5$~[see the inset of Fig.~\ref{fig:PDF}(b)].

It has been analytically shown by \citet{Chertkov:PRL1998} that the passive scalar dissipation has a stretched exponential distribution. This profile is given by $P(\epsilon_T) \sim \beta\exp(- m\epsilon_T^{*\alpha})$ for $\epsilon_T^* \gg 1$. Interestingly, the  PDFs of thermal dissipation for RBC are also stretched exponential for both bulk and boundary layers. Our observation is consistent with earlier studies~\cite{Emran:JFM2008,He:PRE2009,Zhang:JFM2017}. For bulk dissipation, the stretching exponent $\alpha=0.47$ for $\mathrm{Pr}=1$, and $\alpha=0.67$ for $\mathrm{Pr}=100$. The corresponding exponents for the boundary layers are 0.32 and 0.40 respectively. 

\textcolor{black}{Clearly, for both Pr, the tails of the PDFs are stretched more for the boundary layer dissipation. This is expected because extreme events are more frequent in the boundary layers than in the bulk; note that $\epsilon_T(\mathbf{r})$ is stronger in the boundary layers. Further, for both bulk and boundary layer dissipation, $\alpha$'s are smaller for $\mathrm{Pr}=1$. Thus, the tails of the PDFs are stretched more for $\mathrm{Pr=1}$, implying stronger spatial intermittency of thermal dissipation for the lower Pr fluid. This is because for Pr=1, convection is more turbulent than that for Pr=100, causing the temperature fluctuations to be more pronounced for the former.}     

\section{Conclusions} \label{sec:conclusion} 
In this paper, we present scaling relations for thermal dissipation rate in the bulk and in the boundary layers in turbulent convection. Using numerical simulations of RBC, we show that compared to passive scalar turbulence, the thermal dissipation rate in the bulk is suppressed by a factor of $\mathrm{Ra}^{-0.22}$ for $\mathrm{Pr=1}$ and $\mathrm{Ra}^{-0.25}$ for $\mathrm{Pr=100}$. Further, unlike viscous dissipation, the total thermal dissipation in the boundary layers is greater than that in the bulk. The ratio of the boundary layer and the bulk dissipation is roughly constant for $\mathrm{Pr=1}$, and decreases weakly with Ra for $\mathrm{Pr=100}$. 

We also show that the probability distribution functions of thermal dissipation rate, both in the bulk and in the boundary layers, are stretched exponential, similar to passive scalar dissipation. The stretching exponent for the PDFs of boundary layer dissipation is lower than that of bulk dissipation, implying that extreme events occur more often in the boundary layers than in the bulk. We also show that the spatial intermittency of thermal dissipation is stronger for lower Pr fluids. 

The results presented in this paper are important for modelling thermal convection. For example, we may need to incorporate the suppression of thermal dissipation in the bulk in the scaling analysis for Pe and Nu. Thus far, our analysis has been for $\mathrm{Pr} \geq 1$. We need to extend them to low Pr convection for a comprehensive modelling of thermal convection.

\section*{Acknowledgements}
We thank A. Pandey, A. Guha, and R. Samuel for useful discussions. Our numerical simulations were performed on Shaheen II at {\sc Kaust} supercomputing laboratory, Saudi Arabia, under the project k1052.  This work was supported by the research grants PLANEX/PHY/2015239 from Indian Space Research Organisation, India, and by the Department of Science and Technology, India (INT/RUS/RSF/P-03) and Russian Science Foundation Russia (RSF-16-41-02012) for the Indo-Russian project.


\begin{thebibliography}{47}%
\makeatletter
\providecommand \@ifxundefined [1]{%
 \@ifx{#1\undefined}
}%
\providecommand \@ifnum [1]{%
 \ifnum #1\expandafter \@firstoftwo
 \else \expandafter \@secondoftwo
 \fi
}%
\providecommand \@ifx [1]{%
 \ifx #1\expandafter \@firstoftwo
 \else \expandafter \@secondoftwo
 \fi
}%
\providecommand \natexlab [1]{#1}%
\providecommand \enquote  [1]{``#1''}%
\providecommand \bibnamefont  [1]{#1}%
\providecommand \bibfnamefont [1]{#1}%
\providecommand \citenamefont [1]{#1}%
\providecommand \href@noop [0]{\@secondoftwo}%
\providecommand \href [0]{\begingroup \@sanitize@url \@href}%
\providecommand \@href[1]{\@@startlink{#1}\@@href}%
\providecommand \@@href[1]{\endgroup#1\@@endlink}%
\providecommand \@sanitize@url [0]{\catcode `\\12\catcode `\$12\catcode
  `\&12\catcode `\#12\catcode `\^12\catcode `\_12\catcode `\%12\relax}%
\providecommand \@@startlink[1]{}%
\providecommand \@@endlink[0]{}%
\providecommand \url  [0]{\begingroup\@sanitize@url \@url }%
\providecommand \@url [1]{\endgroup\@href {#1}{\urlprefix }}%
\providecommand \urlprefix  [0]{URL }%
\providecommand \Eprint [0]{\href }%
\providecommand \doibase [0]{http://dx.doi.org/}%
\providecommand \selectlanguage [0]{\@gobble}%
\providecommand \bibinfo  [0]{\@secondoftwo}%
\providecommand \bibfield  [0]{\@secondoftwo}%
\providecommand \translation [1]{[#1]}%
\providecommand \BibitemOpen [0]{}%
\providecommand \bibitemStop [0]{}%
\providecommand \bibitemNoStop [0]{.\EOS\space}%
\providecommand \EOS [0]{\spacefactor3000\relax}%
\providecommand \BibitemShut  [1]{\csname bibitem#1\endcsname}%
\let\auto@bib@innerbib\@empty
\bibitem [{\citenamefont {Obukhov}(1949)}]{Obukhov:1949}%
  \BibitemOpen
  \bibfield  {author} {\bibinfo {author} {\bibfnamefont {A.~M.}\ \bibnamefont
  {Obukhov}},\ }\bibfield  {title} {\enquote {\bibinfo {title} {Structure of
  the temperature field in a turbulent flow},}\ }\href@noop {} {\bibfield
  {journal} {\bibinfo  {journal} {Isv. Geogr. Geophys. Ser.}\ }\textbf
  {\bibinfo {volume} {13}},\ \bibinfo {pages} {58--69} (\bibinfo {year}
  {1949})}\BibitemShut {NoStop}%
\bibitem [{\citenamefont {Corrsin}(1951)}]{Corrsin:JAP1951}%
  \BibitemOpen
  \bibfield  {author} {\bibinfo {author} {\bibfnamefont {S.}~\bibnamefont
  {Corrsin}},\ }\bibfield  {title} {\enquote {\bibinfo {title} {{On the
  spectrum of isotropic temperature fluctuations in an isotropic
  turbulence}},}\ }\href@noop {} {\bibfield  {journal} {\bibinfo  {journal} {J.
  Appl. Phys.}\ }\textbf {\bibinfo {volume} {22}},\ \bibinfo {pages} {469--473}
  (\bibinfo {year} {1951})}\BibitemShut {NoStop}%
\bibitem [{\citenamefont {Lesieur}(2008)}]{Lesieur:book:Turbulence}%
  \BibitemOpen
  \bibfield  {author} {\bibinfo {author} {\bibfnamefont {M.}~\bibnamefont
  {Lesieur}},\ }\href@noop {} {\emph {\bibinfo {title} {{Turbulence in
  Fluids}}}}\ (\bibinfo  {publisher} {Springer-Verlag},\ \bibinfo {address}
  {Dordrecht},\ \bibinfo {year} {2008})\BibitemShut {NoStop}%
\bibitem [{\citenamefont {Verma}(2018)}]{Verma:book:BDF}%
  \BibitemOpen
  \bibfield  {author} {\bibinfo {author} {\bibfnamefont {M.~K.}\ \bibnamefont
  {Verma}},\ }\href@noop {} {\emph {\bibinfo {title} {Physics of Buoyant
  Flows}}}\ (\bibinfo  {publisher} {World Scientific},\ \bibinfo {address}
  {Singapore},\ \bibinfo {year} {2018})\BibitemShut {NoStop}%
\bibitem [{\citenamefont {Ahlers}, \citenamefont {Grossmann},\ and\
  \citenamefont {Lohse}(2009)}]{Ahlers:RMP2009}%
  \BibitemOpen
  \bibfield  {author} {\bibinfo {author} {\bibfnamefont {G.}~\bibnamefont
  {Ahlers}}, \bibinfo {author} {\bibfnamefont {S.}~\bibnamefont {Grossmann}}, \
  and\ \bibinfo {author} {\bibfnamefont {D.}~\bibnamefont {Lohse}},\ }\bibfield
   {title} {\enquote {\bibinfo {title} {{Heat transfer and large scale dynamics
  in turbulent Rayleigh-B{\'e}nard convection}},}\ }\href@noop {} {\bibfield
  {journal} {\bibinfo  {journal} {Rev. Mod. Phys.}\ }\textbf {\bibinfo {volume}
  {81}},\ \bibinfo {pages} {503--537} (\bibinfo {year} {2009})}\BibitemShut
  {NoStop}%
\bibitem [{\citenamefont {Lohse}\ and\ \citenamefont
  {Xia}(2010)}]{Lohse:ARFM2010}%
  \BibitemOpen
  \bibfield  {author} {\bibinfo {author} {\bibfnamefont {D.}~\bibnamefont
  {Lohse}}\ and\ \bibinfo {author} {\bibfnamefont {K.-Q.}\ \bibnamefont
  {Xia}},\ }\bibfield  {title} {\enquote {\bibinfo {title} {{Small-scale
  properties of turbulent Rayleigh{\textendash}B{\'e}nard convection}},}\
  }\href@noop {} {\bibfield  {journal} {\bibinfo  {journal} {Annu. Rev. Fluid
  Mech.}\ }\textbf {\bibinfo {volume} {42}},\ \bibinfo {pages} {335--364}
  (\bibinfo {year} {2010})}\BibitemShut {NoStop}%
\bibitem [{\citenamefont {Verma}, \citenamefont {Kumar},\ and\ \citenamefont
  {Pandey}(2017)}]{Verma:NJP2017}%
  \BibitemOpen
  \bibfield  {author} {\bibinfo {author} {\bibfnamefont {M.~K.}\ \bibnamefont
  {Verma}}, \bibinfo {author} {\bibfnamefont {A.}~\bibnamefont {Kumar}}, \ and\
  \bibinfo {author} {\bibfnamefont {A.}~\bibnamefont {Pandey}},\ }\bibfield
  {title} {\enquote {\bibinfo {title} {{Phenomenology of buoyancy-driven
  turbulence: recent results}},}\ }\href@noop {} {\bibfield  {journal}
  {\bibinfo  {journal} {New J. Phys.}\ }\textbf {\bibinfo {volume} {19}},\
  \bibinfo {pages} {025012} (\bibinfo {year} {2017})}\BibitemShut {NoStop}%
\bibitem [{\citenamefont {Shraiman}\ and\ \citenamefont
  {Siggia}(1990)}]{Shraiman:PRA1990}%
  \BibitemOpen
  \bibfield  {author} {\bibinfo {author} {\bibfnamefont {B.~I.}\ \bibnamefont
  {Shraiman}}\ and\ \bibinfo {author} {\bibfnamefont {E.~D.}\ \bibnamefont
  {Siggia}},\ }\bibfield  {title} {\enquote {\bibinfo {title} {{Heat transport
  in high-Rayleigh-number convection}},}\ }\href@noop {} {\bibfield  {journal}
  {\bibinfo  {journal} {Phys. Rev. A}\ }\textbf {\bibinfo {volume} {42}},\
  \bibinfo {pages} {3650--3653} (\bibinfo {year} {1990})}\BibitemShut {NoStop}%
\bibitem [{\citenamefont {Kraichnan}(1962)}]{Kraichnan:PF1962Convection}%
  \BibitemOpen
  \bibfield  {author} {\bibinfo {author} {\bibfnamefont {R.~H.}\ \bibnamefont
  {Kraichnan}},\ }\bibfield  {title} {\enquote {\bibinfo {title} {{Turbulent
  thermal convection at arbitrary prandtl number}},}\ }\href@noop {} {\bibfield
   {journal} {\bibinfo  {journal} {Phys. Fluids}\ }\textbf {\bibinfo {volume}
  {5}},\ \bibinfo {pages} {1374--1389} (\bibinfo {year} {1962})}\BibitemShut
  {NoStop}%
\bibitem [{\citenamefont {Verma}\ \emph {et~al.}(2012)\citenamefont {Verma},
  \citenamefont {Mishra}, \citenamefont {Pandey},\ and\ \citenamefont
  {Paul}}]{Verma:PRE2012}%
  \BibitemOpen
  \bibfield  {author} {\bibinfo {author} {\bibfnamefont {M.~K.}\ \bibnamefont
  {Verma}}, \bibinfo {author} {\bibfnamefont {P.~K.}\ \bibnamefont {Mishra}},
  \bibinfo {author} {\bibfnamefont {A.}~\bibnamefont {Pandey}}, \ and\ \bibinfo
  {author} {\bibfnamefont {S.}~\bibnamefont {Paul}},\ }\bibfield  {title}
  {\enquote {\bibinfo {title} {{Scalings of field correlations and heat
  transport in turbulent convection}},}\ }\href@noop {} {\bibfield  {journal}
  {\bibinfo  {journal} {Phys. Rev. E}\ }\textbf {\bibinfo {volume} {85}},\
  \bibinfo {pages} {016310} (\bibinfo {year} {2012})}\BibitemShut {NoStop}%
\bibitem [{\citenamefont {Grossmann}\ and\ \citenamefont
  {Lohse}(2000)}]{Grossmann:JFM2000}%
  \BibitemOpen
  \bibfield  {author} {\bibinfo {author} {\bibfnamefont {S.}~\bibnamefont
  {Grossmann}}\ and\ \bibinfo {author} {\bibfnamefont {D.}~\bibnamefont
  {Lohse}},\ }\bibfield  {title} {\enquote {\bibinfo {title} {{Scaling in
  thermal convection: a unifying theory}},}\ }\href@noop {} {\bibfield
  {journal} {\bibinfo  {journal} {J. Fluid Mech.}\ }\textbf {\bibinfo {volume}
  {407}},\ \bibinfo {pages} {27--56} (\bibinfo {year} {2000})}\BibitemShut
  {NoStop}%
\bibitem [{\citenamefont {Grossmann}\ and\ \citenamefont
  {Lohse}(2001)}]{Grossmann:PRL2001}%
  \BibitemOpen
  \bibfield  {author} {\bibinfo {author} {\bibfnamefont {S.}~\bibnamefont
  {Grossmann}}\ and\ \bibinfo {author} {\bibfnamefont {D.}~\bibnamefont
  {Lohse}},\ }\bibfield  {title} {\enquote {\bibinfo {title} {{Thermal
  convection for large Prandtl numbers}},}\ }\href@noop {} {\bibfield
  {journal} {\bibinfo  {journal} {Phys. Rev. Lett.}\ }\textbf {\bibinfo
  {volume} {86}},\ \bibinfo {pages} {3316--3319} (\bibinfo {year}
  {2001})}\BibitemShut {NoStop}%
\bibitem [{\citenamefont {Malkus}(1954)}]{Malkus:PRSA1954}%
  \BibitemOpen
  \bibfield  {author} {\bibinfo {author} {\bibfnamefont {W.~V.~R.}\
  \bibnamefont {Malkus}},\ }\bibfield  {title} {\enquote {\bibinfo {title}
  {{The Heat Transport and Spectrum of Thermal Turbulence}},}\ }\href@noop {}
  {\bibfield  {journal} {\bibinfo  {journal} {Proceedings of the Royal Society
  of London. Series A}\ }\textbf {\bibinfo {volume} {225}},\ \bibinfo {pages}
  {196--212} (\bibinfo {year} {1954})}\BibitemShut {NoStop}%
\bibitem [{\citenamefont {Castaing}\ \emph {et~al.}(1989)\citenamefont
  {Castaing}, \citenamefont {Gunaratne}, \citenamefont {{Kadanoff, L. P.}},
  \citenamefont {Libchaber},\ and\ \citenamefont {Heslot}}]{Castaing:JFM1989}%
  \BibitemOpen
  \bibfield  {author} {\bibinfo {author} {\bibfnamefont {B.}~\bibnamefont
  {Castaing}}, \bibinfo {author} {\bibfnamefont {G.}~\bibnamefont {Gunaratne}},
  \bibinfo {author} {\bibnamefont {{Kadanoff, L. P.}}}, \bibinfo {author}
  {\bibfnamefont {A.}~\bibnamefont {Libchaber}}, \ and\ \bibinfo {author}
  {\bibfnamefont {F.}~\bibnamefont {Heslot}},\ }\bibfield  {title} {\enquote
  {\bibinfo {title} {{Scaling of hard thermal turbulence in Rayleigh-B{\'e}nard
  convection}},}\ }\href@noop {} {\bibfield  {journal} {\bibinfo  {journal} {J.
  Fluid Mech.}\ }\textbf {\bibinfo {volume} {204}},\ \bibinfo {pages} {1--30}
  (\bibinfo {year} {1989})}\BibitemShut {NoStop}%
\bibitem [{\citenamefont {Grossmann}\ and\ \citenamefont
  {Lohse}(2002)}]{Grossmann:PRE2002}%
  \BibitemOpen
  \bibfield  {author} {\bibinfo {author} {\bibfnamefont {S.}~\bibnamefont
  {Grossmann}}\ and\ \bibinfo {author} {\bibfnamefont {D.}~\bibnamefont
  {Lohse}},\ }\bibfield  {title} {\enquote {\bibinfo {title} {{Prandtl and
  Rayleigh number dependence of the Reynolds number in turbulent thermal
  convection}},}\ }\href@noop {} {\bibfield  {journal} {\bibinfo  {journal}
  {Phys. Rev. E}\ }\textbf {\bibinfo {volume} {66}},\ \bibinfo {pages} {016305}
  (\bibinfo {year} {2002})}\BibitemShut {NoStop}%
\bibitem [{\citenamefont {Qiu}\ and\ \citenamefont {Tong}(2002)}]{Qiu:PRE2002}%
  \BibitemOpen
  \bibfield  {author} {\bibinfo {author} {\bibfnamefont {X.-L.}\ \bibnamefont
  {Qiu}}\ and\ \bibinfo {author} {\bibfnamefont {P.}~\bibnamefont {Tong}},\
  }\bibfield  {title} {\enquote {\bibinfo {title} {{Temperature oscillations in
  turbulent Rayleigh-B{\'e}nard convection}},}\ }\href@noop {} {\bibfield
  {journal} {\bibinfo  {journal} {Phys. Rev. E}\ }\textbf {\bibinfo {volume}
  {66}},\ \bibinfo {pages} {026308} (\bibinfo {year} {2002})}\BibitemShut
  {NoStop}%
\bibitem [{\citenamefont {Qiu}\ \emph {et~al.}(2004)\citenamefont {Qiu},
  \citenamefont {Shang}, \citenamefont {Tong},\ and\ \citenamefont
  {Xia}}]{Qiu:PF2004}%
  \BibitemOpen
  \bibfield  {author} {\bibinfo {author} {\bibfnamefont {X.-L.}\ \bibnamefont
  {Qiu}}, \bibinfo {author} {\bibfnamefont {X.-D.}\ \bibnamefont {Shang}},
  \bibinfo {author} {\bibfnamefont {P.}~\bibnamefont {Tong}}, \ and\ \bibinfo
  {author} {\bibfnamefont {K.-Q.}\ \bibnamefont {Xia}},\ }\bibfield  {title}
  {\enquote {\bibinfo {title} {{Velocity oscillations in turbulent
  Rayleigh–B\'{e}nard convection}},}\ }\href@noop {} {\bibfield  {journal}
  {\bibinfo  {journal} {Phys. Fluids}\ }\textbf {\bibinfo {volume} {16}},\
  \bibinfo {pages} {412--423} (\bibinfo {year} {2004})}\BibitemShut {NoStop}%
\bibitem [{\citenamefont {Brown}, \citenamefont {Funfschilling},\ and\
  \citenamefont {Ahlers}(2007)}]{Brown:JSM2007}%
  \BibitemOpen
  \bibfield  {author} {\bibinfo {author} {\bibfnamefont {E.}~\bibnamefont
  {Brown}}, \bibinfo {author} {\bibfnamefont {D.}~\bibnamefont
  {Funfschilling}}, \ and\ \bibinfo {author} {\bibfnamefont {G.}~\bibnamefont
  {Ahlers}},\ }\bibfield  {title} {\enquote {\bibinfo {title} {{Anomalous
  Reynolds-number scaling in turbulent Rayleigh{\textendash}B{\'e}nard
  convection}},}\ }\href@noop {} {\bibfield  {journal} {\bibinfo  {journal} {J.
  Stat. Mech. Theor. Exp.}\ }\textbf {\bibinfo {volume} {2007}},\ \bibinfo
  {pages} {P10005} (\bibinfo {year} {2007})}\BibitemShut {NoStop}%
\bibitem [{\citenamefont {Funfschilling}\ \emph {et~al.}(2005)\citenamefont
  {Funfschilling}, \citenamefont {Brown}, \citenamefont {Nikolaenko},\ and\
  \citenamefont {Ahlers}}]{Funfschilling:JFM2005}%
  \BibitemOpen
  \bibfield  {author} {\bibinfo {author} {\bibfnamefont {D.}~\bibnamefont
  {Funfschilling}}, \bibinfo {author} {\bibfnamefont {E.}~\bibnamefont
  {Brown}}, \bibinfo {author} {\bibfnamefont {A.}~\bibnamefont {Nikolaenko}}, \
  and\ \bibinfo {author} {\bibfnamefont {G.}~\bibnamefont {Ahlers}},\
  }\bibfield  {title} {\enquote {\bibinfo {title} {{Heat transport by turbulent
  Rayleigh-B\'{e}nard convection in cylindrical samples with aspect ratio one
  and larger}},}\ }\href@noop {} {\bibfield  {journal} {\bibinfo  {journal} {J.
  Fluid Mech.}\ }\textbf {\bibinfo {volume} {536}},\ \bibinfo {pages}
  {145--154} (\bibinfo {year} {2005})}\BibitemShut {NoStop}%
\bibitem [{\citenamefont {Nikolaenko}\ \emph {et~al.}(2005)\citenamefont
  {Nikolaenko}, \citenamefont {Brown}, \citenamefont {Funfschilling},\ and\
  \citenamefont {Ahlers}}]{Nikolaenko:JFM2005}%
  \BibitemOpen
  \bibfield  {author} {\bibinfo {author} {\bibfnamefont {A.}~\bibnamefont
  {Nikolaenko}}, \bibinfo {author} {\bibfnamefont {E.}~\bibnamefont {Brown}},
  \bibinfo {author} {\bibfnamefont {D.}~\bibnamefont {Funfschilling}}, \ and\
  \bibinfo {author} {\bibfnamefont {G.}~\bibnamefont {Ahlers}},\ }\bibfield
  {title} {\enquote {\bibinfo {title} {{Heat transport by turbulent
  Rayleigh-B{\'e}nard convection in cylindrical cells with aspect ratio one and
  less}},}\ }\href@noop {} {\bibfield  {journal} {\bibinfo  {journal} {J. Fluid
  Mech.}\ }\textbf {\bibinfo {volume} {523}},\ \bibinfo {pages} {251--260}
  (\bibinfo {year} {2005})}\BibitemShut {NoStop}%
\bibitem [{\citenamefont {He}\ \emph {et~al.}(2012)\citenamefont {He},
  \citenamefont {Funfschilling}, \citenamefont {Bodenschadtz},\ and\
  \citenamefont {Ahlers}}]{He:NJP2012}%
  \BibitemOpen
  \bibfield  {author} {\bibinfo {author} {\bibfnamefont {X.}~\bibnamefont
  {He}}, \bibinfo {author} {\bibfnamefont {D.}~\bibnamefont {Funfschilling}},
  \bibinfo {author} {\bibfnamefont {E.}~\bibnamefont {Bodenschadtz}}, \ and\
  \bibinfo {author} {\bibfnamefont {G.}~\bibnamefont {Ahlers}},\ }\bibfield
  {title} {\enquote {\bibinfo {title} {{Heat transport by turbulent
  Rayleigh-B\'{e}nard convection for Pr $\simeq 0.8$ and $4 \times 10^{11}
  \lesssim$ Ra $\lesssim 2 \times 10^{14}$: ultimate-state transition for
  aspect ratio $\Gamma=1.00$}},}\ }\href@noop {} {\bibfield  {journal}
  {\bibinfo  {journal} {New J. Phys.}\ }\textbf {\bibinfo {volume} {14}},\
  \bibinfo {pages} {063030} (\bibinfo {year} {2012})}\BibitemShut {NoStop}%
\bibitem [{\citenamefont {Ahlers}\ \emph {et~al.}(2012)\citenamefont {Ahlers},
  \citenamefont {He}, \citenamefont {Funfschilling},\ and\ \citenamefont
  {Bodenschadtz}}]{Ahlers:NJP2012}%
  \BibitemOpen
  \bibfield  {author} {\bibinfo {author} {\bibfnamefont {G.}~\bibnamefont
  {Ahlers}}, \bibinfo {author} {\bibfnamefont {X.}~\bibnamefont {He}}, \bibinfo
  {author} {\bibfnamefont {D.}~\bibnamefont {Funfschilling}}, \ and\ \bibinfo
  {author} {\bibfnamefont {E.}~\bibnamefont {Bodenschadtz}},\ }\bibfield
  {title} {\enquote {\bibinfo {title} {{Heat transport by turbulent
  Rayleigh-B\'{e}nard convection for Pr $\simeq 0.8$ and $3 \times 10^{12}
  \lesssim$ Ra $\lesssim 10^{15}$: Aspect ratio $\Gamma=0.50$}},}\ }\href@noop
  {} {\bibfield  {journal} {\bibinfo  {journal} {New J. Phys.}\ }\textbf
  {\bibinfo {volume} {14}},\ \bibinfo {pages} {103012} (\bibinfo {year}
  {2012})}\BibitemShut {NoStop}%
\bibitem [{\citenamefont {Vial}\ and\ \citenamefont
  {Hern{\'a}ndes}(2017)}]{Vial:PF2017}%
  \BibitemOpen
  \bibfield  {author} {\bibinfo {author} {\bibfnamefont {M.}~\bibnamefont
  {Vial}}\ and\ \bibinfo {author} {\bibfnamefont {R.~H.}\ \bibnamefont
  {Hern{\'a}ndes}},\ }\bibfield  {title} {\enquote {\bibinfo {title} {{Feedback
  control and heat transfer measurements in a Rayleigh-B{\'e}nard convection
  cell}},}\ }\href@noop {} {\bibfield  {journal} {\bibinfo  {journal} {Phys.
  Fluids}\ }\textbf {\bibinfo {volume} {29}},\ \bibinfo {pages} {074103}
  (\bibinfo {year} {2017})}\BibitemShut {NoStop}%
\bibitem [{\citenamefont {\color{black}Verzicco}\ and\ \citenamefont
  {Camussi}(2003)}]{Verzicco:JFM2003}%
  \BibitemOpen
  \bibfield  {author} {\bibinfo {author} {\bibfnamefont {R.}~\bibnamefont
  {\color{black}Verzicco}}\ and\ \bibinfo {author} {\bibfnamefont
  {R.}~\bibnamefont {Camussi}},\ }\bibfield  {title} {\enquote {\bibinfo
  {title} {{Numerical experiments on strongly turbulent thermal convection in a
  slender cylindrical cell}},}\ }\href@noop {} {\bibfield  {journal} {\bibinfo
  {journal} {J. Fluid Mech.}\ }\textbf {\bibinfo {volume} {477}},\ \bibinfo
  {pages} {19--49} (\bibinfo {year} {2003})}\BibitemShut {NoStop}%
\bibitem [{\citenamefont {Scheel}, \citenamefont {Kim},\ and\ \citenamefont
  {White}(2012)}]{Scheel:JFM2012}%
  \BibitemOpen
  \bibfield  {author} {\bibinfo {author} {\bibfnamefont {J.~D.}\ \bibnamefont
  {Scheel}}, \bibinfo {author} {\bibfnamefont {E.}~\bibnamefont {Kim}}, \ and\
  \bibinfo {author} {\bibfnamefont {K.~R.}\ \bibnamefont {White}},\ }\bibfield
  {title} {\enquote {\bibinfo {title} {{Thermal and viscous boundary layers in
  turbulent Rayleigh{\textendash}B{\'e}nard convection}},}\ }\href@noop {}
  {\bibfield  {journal} {\bibinfo  {journal} {J. Fluid Mech.}\ }\textbf
  {\bibinfo {volume} {711}},\ \bibinfo {pages} {281--305} (\bibinfo {year}
  {2012})}\BibitemShut {NoStop}%
\bibitem [{\citenamefont {Scheel}\ and\ \citenamefont
  {Schumacher}(2014)}]{Scheel:JFM2014}%
  \BibitemOpen
  \bibfield  {author} {\bibinfo {author} {\bibfnamefont {J.~D.}\ \bibnamefont
  {Scheel}}\ and\ \bibinfo {author} {\bibfnamefont {J.}~\bibnamefont
  {Schumacher}},\ }\bibfield  {title} {\enquote {\bibinfo {title} {{Local
  boundary layer scales in turbulent Rayleigh{\textendash}B{\'e}nard
  convection}},}\ }\href@noop {} {\bibfield  {journal} {\bibinfo  {journal} {J.
  Fluid Mech.}\ }\textbf {\bibinfo {volume} {758}},\ \bibinfo {pages}
  {344--373} (\bibinfo {year} {2014})}\BibitemShut {NoStop}%
\bibitem [{\citenamefont {Waleffe}, \citenamefont {Boonkasame},\ and\
  \citenamefont {Smith}(2015)}]{Waleffe:PF2015}%
  \BibitemOpen
  \bibfield  {author} {\bibinfo {author} {\bibfnamefont {F.}~\bibnamefont
  {Waleffe}}, \bibinfo {author} {\bibfnamefont {A.}~\bibnamefont {Boonkasame}},
  \ and\ \bibinfo {author} {\bibfnamefont {L.~M.}\ \bibnamefont {Smith}},\
  }\bibfield  {title} {\enquote {\bibinfo {title} {{Heat transport by coherent
  Rayleigh-B{\'e}nard convection}},}\ }\href@noop {} {\bibfield  {journal}
  {\bibinfo  {journal} {Phys. Fluids}\ }\textbf {\bibinfo {volume} {27}},\
  \bibinfo {pages} {051702} (\bibinfo {year} {2015})}\BibitemShut {NoStop}%
\bibitem [{\citenamefont {Verma}, \citenamefont {Ambhire},\ and\ \citenamefont
  {Pandey}(2015)}]{Verma:PF2015Reversal}%
  \BibitemOpen
  \bibfield  {author} {\bibinfo {author} {\bibfnamefont {M.~K.}\ \bibnamefont
  {Verma}}, \bibinfo {author} {\bibfnamefont {S.~C.}\ \bibnamefont {Ambhire}},
  \ and\ \bibinfo {author} {\bibfnamefont {A.}~\bibnamefont {Pandey}},\
  }\bibfield  {title} {\enquote {\bibinfo {title} {{Flow reversals in turbulent
  convection with free-slip walls}},}\ }\href@noop {} {\bibfield  {journal}
  {\bibinfo  {journal} {Phys. Fluids}\ }\textbf {\bibinfo {volume} {27}},\
  \bibinfo {pages} {047102} (\bibinfo {year} {2015})}\BibitemShut {NoStop}%
\bibitem [{\citenamefont {Zhou}\ and\ \citenamefont
  {Chen}(2018)}]{Zhou:PF2018}%
  \BibitemOpen
  \bibfield  {author} {\bibinfo {author} {\bibfnamefont {W.-F.}\ \bibnamefont
  {Zhou}}\ and\ \bibinfo {author} {\bibfnamefont {J.}~\bibnamefont {Chen}},\
  }\bibfield  {title} {\enquote {\bibinfo {title} {{Letter: Similarity model
  for corner roll in turbulent Rayleigh-B{\'e}nard convection}},}\ }\href@noop
  {} {\bibfield  {journal} {\bibinfo  {journal} {Phys. Fluids}\ }\textbf
  {\bibinfo {volume} {30}},\ \bibinfo {pages} {111705} (\bibinfo {year}
  {2018})}\BibitemShut {NoStop}%
\bibitem [{\citenamefont {\color{black}Pandey}\ and\ \citenamefont
  {Verma}(2016)}]{Pandey:PF2016}%
  \BibitemOpen
  \bibfield  {author} {\bibinfo {author} {\bibfnamefont {A.}~\bibnamefont
  {\color{black}Pandey}}\ and\ \bibinfo {author} {\bibfnamefont {M.~K.}\
  \bibnamefont {Verma}},\ }\bibfield  {title} {\enquote {\bibinfo {title}
  {{Scaling of large-scale quantities in Rayleigh-B{\'e}nard convection}},}\
  }\href@noop {} {\bibfield  {journal} {\bibinfo  {journal} {Phys. Fluids}\
  }\textbf {\bibinfo {volume} {28}},\ \bibinfo {pages} {095105} (\bibinfo
  {year} {2016})}\BibitemShut {NoStop}%
\bibitem [{\citenamefont {Pandey}\ \emph {et~al.}(2016)\citenamefont {Pandey},
  \citenamefont {Kumar}, \citenamefont {Chatterjee},\ and\ \citenamefont
  {Verma}}]{Pandey:PRE2016}%
  \BibitemOpen
  \bibfield  {author} {\bibinfo {author} {\bibfnamefont {A.}~\bibnamefont
  {Pandey}}, \bibinfo {author} {\bibfnamefont {A.}~\bibnamefont {Kumar}},
  \bibinfo {author} {\bibfnamefont {A.~G.}\ \bibnamefont {Chatterjee}}, \ and\
  \bibinfo {author} {\bibfnamefont {M.~K.}\ \bibnamefont {Verma}},\ }\bibfield
  {title} {\enquote {\bibinfo {title} {{Dynamics of large-scale quantities in
  Rayleigh-B{\'e}nard convection}},}\ }\href@noop {} {\bibfield  {journal}
  {\bibinfo  {journal} {Phys. Rev. E}\ }\textbf {\bibinfo {volume} {94}},\
  \bibinfo {pages} {053106} (\bibinfo {year} {2016})}\BibitemShut {NoStop}%
\bibitem [{\citenamefont {Puthenveettil}\ and\ \citenamefont
  {Arakeri}(2005)}]{Baburaj:JFM2005}%
  \BibitemOpen
  \bibfield  {author} {\bibinfo {author} {\bibfnamefont {B.~A.}\ \bibnamefont
  {Puthenveettil}}\ and\ \bibinfo {author} {\bibfnamefont {J.~H.}\ \bibnamefont
  {Arakeri}},\ }\bibfield  {title} {\enquote {\bibinfo {title} {{Plume
  structure in high-Rayleigh-number convection}},}\ }\href@noop {} {\bibfield
  {journal} {\bibinfo  {journal} {J. Fluid Mech.}\ }\textbf {\bibinfo {volume}
  {542}},\ \bibinfo {pages} {217--249} (\bibinfo {year} {2005})}\BibitemShut
  {NoStop}%
\bibitem [{\citenamefont {Puthenveettil}, \citenamefont {Ananthakrishna},\ and\
  \citenamefont {Arakeri}(2005)}]{Baburaj:JFM2005a}%
  \BibitemOpen
  \bibfield  {author} {\bibinfo {author} {\bibfnamefont {B.~A.}\ \bibnamefont
  {Puthenveettil}}, \bibinfo {author} {\bibfnamefont {G.}~\bibnamefont
  {Ananthakrishna}}, \ and\ \bibinfo {author} {\bibfnamefont {J.~H.}\
  \bibnamefont {Arakeri}},\ }\bibfield  {title} {\enquote {\bibinfo {title}
  {{The multifractal nature of plume structure in high-Rayleigh-number
  convection}},}\ }\href@noop {} {\bibfield  {journal} {\bibinfo  {journal} {J.
  Fluid Mech.}\ }\textbf {\bibinfo {volume} {526}},\ \bibinfo {pages}
  {245--256} (\bibinfo {year} {2005})}\BibitemShut {NoStop}%
\bibitem [{\citenamefont {\color{black}Bhattacharya}\ \emph
  {et~al.}(2018)\citenamefont {\color{black}Bhattacharya}, \citenamefont
  {Pandey}, \citenamefont {Kumar},\ and\ \citenamefont
  {Verma}}]{Bhattacharya:PF2018}%
  \BibitemOpen
  \bibfield  {author} {\bibinfo {author} {\bibfnamefont {S.}~\bibnamefont
  {\color{black}Bhattacharya}}, \bibinfo {author} {\bibfnamefont
  {A.}~\bibnamefont {Pandey}}, \bibinfo {author} {\bibfnamefont
  {A.}~\bibnamefont {Kumar}}, \ and\ \bibinfo {author} {\bibfnamefont {M.~K.}\
  \bibnamefont {Verma}},\ }\bibfield  {title} {\enquote {\bibinfo {title}
  {{Complexity of viscous dissipation in turbulent thermal convection}},}\
  }\href@noop {} {\bibfield  {journal} {\bibinfo  {journal} {Phys. Fluids}\
  }\textbf {\bibinfo {volume} {30}},\ \bibinfo {pages} {031702} (\bibinfo
  {year} {2018})}\BibitemShut {NoStop}%
\bibitem [{\citenamefont {Zhang}, \citenamefont {Zhou},\ and\ \citenamefont
  {Sun}(2017)}]{Zhang:JFM2017}%
  \BibitemOpen
  \bibfield  {author} {\bibinfo {author} {\bibfnamefont {Y.}~\bibnamefont
  {Zhang}}, \bibinfo {author} {\bibfnamefont {Q.}~\bibnamefont {Zhou}}, \ and\
  \bibinfo {author} {\bibfnamefont {C.}~\bibnamefont {Sun}},\ }\bibfield
  {title} {\enquote {\bibinfo {title} {{Statistics of kinetic and thermal
  energy dissipation rates in two-dimensional turbulent
  Rayleigh{\textendash}B{\'e}nard convection}},}\ }\href@noop {} {\bibfield
  {journal} {\bibinfo  {journal} {J. Fluid Mech.}\ }\textbf {\bibinfo {volume}
  {814}},\ \bibinfo {pages} {165--184} (\bibinfo {year} {2017})}\BibitemShut
  {NoStop}%
\bibitem [{\citenamefont
  {Chandrasekhar}(2013)}]{Chandrasekhar:book:Instability}%
  \BibitemOpen
  \bibfield  {author} {\bibinfo {author} {\bibfnamefont {S.}~\bibnamefont
  {Chandrasekhar}},\ }\href@noop {} {\emph {\bibinfo {title} {{Hydrodynamic and
  Hydromagnetic Stability}}}}\ (\bibinfo  {publisher} {Oxford University
  Press},\ \bibinfo {address} {Oxford},\ \bibinfo {year} {2013})\BibitemShut
  {NoStop}%
\bibitem [{\citenamefont {Kumar}\ and\ \citenamefont
  {Verma}(2018)}]{Kumar:RSOS2018}%
  \BibitemOpen
  \bibfield  {author} {\bibinfo {author} {\bibfnamefont {A.}~\bibnamefont
  {Kumar}}\ and\ \bibinfo {author} {\bibfnamefont {M.~K.}\ \bibnamefont
  {Verma}},\ }\bibfield  {title} {\enquote {\bibinfo {title} {{Applicability of
  Taylor{\textquoteright}s hypothesis in thermally driven turbulence}},}\
  }\href@noop {} {\bibfield  {journal} {\bibinfo  {journal} {Royal Society Open
  Science}\ }\textbf {\bibinfo {volume} {5}},\ \bibinfo {pages} {172152}
  (\bibinfo {year} {2018})}\BibitemShut {NoStop}%
\bibitem [{\citenamefont {Jasak}\ \emph {et~al.}(2007)\citenamefont {Jasak},
  \citenamefont {Jemcov}, \citenamefont {Tukovic} \emph
  {et~al.}}]{Jasak:CD2007}%
  \BibitemOpen
  \bibfield  {author} {\bibinfo {author} {\bibfnamefont {H.}~\bibnamefont
  {Jasak}}, \bibinfo {author} {\bibfnamefont {A.}~\bibnamefont {Jemcov}},
  \bibinfo {author} {\bibfnamefont {Z.}~\bibnamefont {Tukovic}},  \emph
  {et~al.},\ }\bibfield  {title} {\enquote {\bibinfo {title} {{OpenFOAM: A C++
  library for complex physics simulations}},}\ }in\ \href@noop {} {\emph
  {\bibinfo {booktitle} {International workshop on coupled methods in numerical
  dynamics}}},\ Vol.\ \bibinfo {volume} {1000}\ (\bibinfo {organization} {IUC
  Dubrovnik, Croatia},\ \bibinfo {year} {2007})\ pp.\ \bibinfo {pages}
  {1--20}\BibitemShut {NoStop}%
\bibitem [{\citenamefont {Gr{\"o}tzbach}(1983)}]{Grotzbach:JCP1983}%
  \BibitemOpen
  \bibfield  {author} {\bibinfo {author} {\bibfnamefont {G.}~\bibnamefont
  {Gr{\"o}tzbach}},\ }\bibfield  {title} {\enquote {\bibinfo {title} {{Spatial
  resolution requirements for direct numerical simulation of the
  Rayleigh-Bénard convection}},}\ }\href@noop {} {\bibfield  {journal}
  {\bibinfo  {journal} {J. Comput. Phys.}\ }\textbf {\bibinfo {volume} {49}},\
  \bibinfo {pages} {241--264} (\bibinfo {year} {1983})}\BibitemShut {NoStop}%
\bibitem [{\citenamefont {Shi}, \citenamefont {Emran},\ and\ \citenamefont
  {Schumacher}(2012)}]{Shi:JFM2012}%
  \BibitemOpen
  \bibfield  {author} {\bibinfo {author} {\bibfnamefont {N.}~\bibnamefont
  {Shi}}, \bibinfo {author} {\bibfnamefont {M.~S.}\ \bibnamefont {Emran}}, \
  and\ \bibinfo {author} {\bibfnamefont {J.}~\bibnamefont {Schumacher}},\
  }\bibfield  {title} {\enquote {\bibinfo {title} {{Boundary layer structure in
  turbulent Rayleigh{\textendash}B{\'e}nard convection}},}\ }\href@noop {}
  {\bibfield  {journal} {\bibinfo  {journal} {J. Fluid Mech.}\ }\textbf
  {\bibinfo {volume} {706}},\ \bibinfo {pages} {5--33} (\bibinfo {year}
  {2012})}\BibitemShut {NoStop}%
\bibitem [{\citenamefont {Batchelor}(1959)}]{Batchelor:JFM1959_largeSc}%
  \BibitemOpen
  \bibfield  {author} {\bibinfo {author} {\bibfnamefont {G.~K.}\ \bibnamefont
  {Batchelor}},\ }\bibfield  {title} {\enquote {\bibinfo {title} {{Small-scale
  variation of convected quantities like temperature in turbulent fluid Part 1.
  General discussion and the case of small conductivity}},}\ }\href@noop {}
  {\bibfield  {journal} {\bibinfo  {journal} {J. Fluid Mech.}\ }\textbf
  {\bibinfo {volume} {5}},\ \bibinfo {pages} {113--133} (\bibinfo {year}
  {1959})}\BibitemShut {NoStop}%
\bibitem [{\citenamefont {Wagner}\ and\ \citenamefont
  {Shishkina}(2013)}]{Wagner:PF2013}%
  \BibitemOpen
  \bibfield  {author} {\bibinfo {author} {\bibfnamefont {S.}~\bibnamefont
  {Wagner}}\ and\ \bibinfo {author} {\bibfnamefont {O.}~\bibnamefont
  {Shishkina}},\ }\bibfield  {title} {\enquote {\bibinfo {title} {{Aspect-ratio
  dependency of Rayleigh-B{\'e}nard convection in box-shaped containers}},}\
  }\href@noop {} {\bibfield  {journal} {\bibinfo  {journal} {Phys. Fluids}\
  }\textbf {\bibinfo {volume} {25}},\ \bibinfo {pages} {085110} (\bibinfo
  {year} {2013})}\BibitemShut {NoStop}%
\bibitem [{\citenamefont {Pandey}, \citenamefont {Verma},\ and\ \citenamefont
  {Mishra}(2014)}]{Pandey:PRE2014}%
  \BibitemOpen
  \bibfield  {author} {\bibinfo {author} {\bibfnamefont {A.}~\bibnamefont
  {Pandey}}, \bibinfo {author} {\bibfnamefont {M.~K.}\ \bibnamefont {Verma}}, \
  and\ \bibinfo {author} {\bibfnamefont {P.~K.}\ \bibnamefont {Mishra}},\
  }\bibfield  {title} {\enquote {\bibinfo {title} {{Scaling of heat flux and
  energy spectrum for very large Prandtl number convection}},}\ }\href@noop {}
  {\bibfield  {journal} {\bibinfo  {journal} {Phys. Rev. E}\ }\textbf {\bibinfo
  {volume} {89}},\ \bibinfo {pages} {023006} (\bibinfo {year}
  {2014})}\BibitemShut {NoStop}%
\bibitem [{\citenamefont {Emran}\ and\ \citenamefont
  {Schumacher}(2008)}]{Emran:JFM2008}%
  \BibitemOpen
  \bibfield  {author} {\bibinfo {author} {\bibfnamefont {M.~S.}\ \bibnamefont
  {Emran}}\ and\ \bibinfo {author} {\bibfnamefont {J.}~\bibnamefont
  {Schumacher}},\ }\bibfield  {title} {\enquote {\bibinfo {title} {{Fine-scale
  statistics of temperature and its derivatives in convective turbulence}},}\
  }\href@noop {} {\bibfield  {journal} {\bibinfo  {journal} {J. Fluid Mech.}\
  }\textbf {\bibinfo {volume} {611}},\ \bibinfo {pages} {13--34} (\bibinfo
  {year} {2008})}\BibitemShut {NoStop}%
\bibitem [{\citenamefont {Shishkina}\ \emph {et~al.}(2017)\citenamefont
  {Shishkina}, \citenamefont {Emran}, \citenamefont {Grossmann},\ and\
  \citenamefont {Lohse}}]{Shishkina:PRF2017}%
  \BibitemOpen
  \bibfield  {author} {\bibinfo {author} {\bibfnamefont {O.}~\bibnamefont
  {Shishkina}}, \bibinfo {author} {\bibfnamefont {M.~S.}\ \bibnamefont
  {Emran}}, \bibinfo {author} {\bibfnamefont {S.}~\bibnamefont {Grossmann}}, \
  and\ \bibinfo {author} {\bibfnamefont {D.}~\bibnamefont {Lohse}},\ }\bibfield
   {title} {\enquote {\bibinfo {title} {{Scaling relations in
  large-Prandtl-number natural thermal convection}},}\ }\href@noop {}
  {\bibfield  {journal} {\bibinfo  {journal} {Phys. Rev. Fluids}\ }\textbf
  {\bibinfo {volume} {2}},\ \bibinfo {pages} {103502} (\bibinfo {year}
  {2017})}\BibitemShut {NoStop}%
\bibitem [{\citenamefont {Chertkov}, \citenamefont {Falkovich},\ and\
  \citenamefont {Kolokolov}(1998)}]{Chertkov:PRL1998}%
  \BibitemOpen
  \bibfield  {author} {\bibinfo {author} {\bibfnamefont {M.}~\bibnamefont
  {Chertkov}}, \bibinfo {author} {\bibfnamefont {G.}~\bibnamefont {Falkovich}},
  \ and\ \bibinfo {author} {\bibfnamefont {I.}~\bibnamefont {Kolokolov}},\
  }\bibfield  {title} {\enquote {\bibinfo {title} {{Intermittent Dissipation of
  a Passive Scalar in Turbulence}},}\ }\href@noop {} {\bibfield  {journal}
  {\bibinfo  {journal} {Phys. Rev. Lett.}\ }\textbf {\bibinfo {volume} {80}},\
  \bibinfo {pages} {2121--2124} (\bibinfo {year} {1998})}\BibitemShut {NoStop}%
\bibitem [{\citenamefont {He}\ and\ \citenamefont {Tong}(2009)}]{He:PRE2009}%
  \BibitemOpen
  \bibfield  {author} {\bibinfo {author} {\bibfnamefont {X.}~\bibnamefont
  {He}}\ and\ \bibinfo {author} {\bibfnamefont {P.}~\bibnamefont {Tong}},\
  }\bibfield  {title} {\enquote {\bibinfo {title} {{Measurements of the thermal
  dissipation field in turbulent Rayleigh-B{\'e}nard convection}},}\
  }\href@noop {} {\bibfield  {journal} {\bibinfo  {journal} {Phys. Rev. E}\
  }\textbf {\bibinfo {volume} {79}},\ \bibinfo {pages} {026306} (\bibinfo
  {year} {2009})}\BibitemShut {NoStop}%
\end{thebibliography}

%

\end{document}